\documentclass[fleqn,usenatbib]{mnras}

\usepackage{newtxtext,newtxmath}

\usepackage[T1]{fontenc}

\DeclareRobustCommand{\VAN}[3]{#2}
\let\VANthebibliography\thebibliography
\def\thebibliography{\DeclareRobustCommand{\VAN}[3]{##3}\VANthebibliography}

\usepackage{graphicx}	
\usepackage{amsmath}	
\usepackage{amssymb}	

\title[LRG-BEASTS: Transmission Spectroscopy of WASP-21b]{LRG-BEASTS: Ground-based Detection of Sodium and a Steep Optical Slope in the Atmosphere of the Highly Inflated Hot-Saturn WASP-21b}

\author[L. Alderson et al.]{
L. Alderson,$^{1,2}$ \thanks{E-mail: lili.alderson@cfa.harvard.edu (LA)}
J. Kirk,$^{2}$ \thanks{E-mail: james.kirk@cfa.harvard.edu (JK)}
M. L\'opez-Morales, $^{2}$
P. J. Wheatley, $^{3,4}$
I. Skillen, $^{5}$ \newauthor
G. W. Henry, $^{6}$ 
C. McGruder, $^{2}$
M. Brogi, $^{3,4,7}$
T. Louden $^{3}$
and G. King $^{3}$
\\
$^{1}$Department of Physics and Astronomy, University of Southampton, Southampton SO17 1BJ, UK\\
$^{2}$Center for Astrophysics $|$ Harvard \& Smithsonian, Cambridge, MA 02138, USA\\
$^{3}$Department of Physics, University of Warwick, Coventry, CV4 7AL, UK\\
$^{4}$Centre for Exoplanets and Habitability, University of Warwick, Gibbet Hill Road, Coventry CV4 7AL, UK \\
$^{5}$Isaac Newton Group of Telescopes, Apartado de Correos 321, 38700 Santa Cruz de la Palma, Spain \\
$^{6}$Center of Excellence in Information Systems, Tennessee State University, Nashville, TN  37209  USA \\
$^{7}$ INAF - Osservatorio Astrofisico di Torino, Via Osservatorio 20, I-10025 Pino Torinese, Italy}

\date{Accepted XXX. Received YYY; in original form ZZZ}

\pubyear{2020}

\begin{document}
\label{firstpage}
\pagerange{\pageref{firstpage}--\pageref{lastpage}}
\maketitle

\begin{abstract}


We present the optical transmission spectrum of the highly inflated Saturn-mass exoplanet WASP-21b, using three transits obtained with the ACAM instrument on the William Herschel Telescope through the LRG-BEASTS survey (Low Resolution Ground-Based Exoplanet Atmosphere Survey using Transmission Spectroscopy). Our transmission spectrum covers a wavelength range of 4635-9000\,\AA, achieving an average transit depth precision of 197\,ppm compared to one atmospheric scale height at 246\,ppm. We detect \ion{Na}{i} absorption in a bin width of 30\,\AA\, at >4$\sigma$ confidence, which extends over 100\,\AA. We see no evidence of absorption from \ion{K}{i}. Atmospheric retrieval analysis of the scattering slope indicates it is too steep for Rayleigh scattering from H$_2$, but is very similar to that of HD\,189733b. The features observed in our transmission spectrum cannot be caused by stellar activity alone, with photometric monitoring of WASP-21 showing it to be an inactive star. We therefore conclude that aerosols in the atmosphere of WASP-21b are giving rise to the steep slope that we observe, and that WASP-21b is an excellent target for infra-red observations to constrain its atmospheric metallicity.

\end{abstract}




\section{Introduction} \label{section:info}

Since the first detection of the atmosphere of an exoplanet \citep{Charbonneau2002}, the atmospheres of $\sim$\,40 exoplanets have been characterised through transmission spectroscopy \citep{Madhusudhan2019}. Such studies have shown that exoplanets have diverse atmospheres, ranging from cloudy with muted atomic and molecular features, to clear and feature rich \citep[e.g.][]{Sing2013, Kreidberg2014, Sing2016, Nikolov2018, Wakeford2018, Carter2019, Spake2019}. Recently, comparative studies of exoplanet atmospheres have revealed that atmospheres appear to transition from clear to cloudy \citep[e.g.][]{Sing2016, Heng2016, Crossfield2017,Pinhas2019}, however, truly clear atmospheres appear to be few and far between \citep{Wakeford2019}.

Early modelling predicted strong, broad absorption features from neutral sodium (\ion{Na}{i}) and potassium (\ion{K}{i}) at wavelengths of 5893\,\AA\, and 7682\,\AA\, respectively \citep{Seager2000, Brown2001}, and while the narrow cores of these features have been detected in many atmospheres at both low \citep[e.g.][]{Charbonneau2002,Huitson2012,Sing2012,Nikolov2013,Wilson2015,Chen2018,Carter2019} and high resolution \cite[e.g.][]{Snellen2008,Redfield2008,Wyttenbach2015,Louden2015,Casasayas2017,Zak2019,Keles2019,Chen2020}, the pressure broadened wings are less commonly observed \citep[e.g.][]{Fischer2016a, Nikolov2018}. This is due to the fact that condensate clouds and photochemical hazes within atmospheres can mute the extended wings or even entirely mask such absorption features \citep[e.g.][]{Pont2008, Sing2015, Kirk2017}. However, clouds and hazes can also lead to blueward scattering slopes \citep[e.g.][]{LecavelierDesEtangs2008,  LecavelierDesEtangs2008a, Sing2011a, Kirk2017, Mallonn2017}, with steeper gradients than predicted from H$_{2}$ scattering alone \citep{Pont2013, Wakeford2015, Pinhas2017}. In order to accurately interpret an exoplanet's transmission spectrum, activity on the host star must also be accounted for, as spots and faculae can mimic absorption signatures within a planet's atmosphere \citep[e.g.][]{Oshagh2014,McCullough2014,Kirk2016,Rackham2017, Cauley2018}. 

In order to better understand the processes driving the observed diversity of exoplanet atmospheres, we need to expand the sample of observed exoplanets. In particular, given the ability for clouds and hazes to mute atmospheric features, it would be beneficial to be able to predict whether an exoplanet will have a clear or cloudy atmosphere given system parameters such as its equilibrium temperature, gravity or stellar metallicity. Tentative evidence suggests that hotter planets are more likely to be cloud free than their cooler counterparts \citep[e.g.][]{Stevenson2016, Heng2016, Fu2017, Crossfield2017, Evans2018, Hoeijmakers2018, Hoeijmakers2019}, however it is possible that clouds \citep{Wakeford2015,Wakeford2017} and hazes \citep{Zahnle2009} may persist in even the hottest atmospheres. 

Building the sample of characterised exoplanet atmospheres at optical wavelengths is the primary aim of the Low-Resolution Ground-Based Exoplanet Atmosphere Survey using Transmission Spectroscopy (LRG-BEASTS; `large beasts'). Optical data is essential to accurately constrain abundances derived from infra-red data \citep[e.g.][]{Benneke2012, Heng2017,Wakeford2018, Pinhas2019}, and therefore such measurements are key to improving our understanding of exoplanet atmospheres. LRG-BEASTS has previously demonstrated that 4-metre class telescopes can obtain transmission spectra with precisions comparable to that of 8 and 10-metre class telescopes. To date, LRG-BEASTS has revealed a haze-induced Rayleigh scattering slope in the atmosphere of HAT-P-18b \citep{Kirk2017}, a grey cloud deck in the atmosphere of WASP-52b \citep{Kirk2016, Louden2017}, a haze in the atmosphere of WASP-80b \citep{Kirk2018} and analysed the supersolar metallicity of the atmosphere of WASP-39b \citep{Kirk2019}.

In this paper we focus on WASP-21b \citep{Bouchy2010}, a 0.30\,$\mathrm{M_{Jup}}$, 1.3\,$\mathrm{R_{Jup}}$ hot Saturn-mass exoplanet \citep{Southworth2012}. WASP-21b orbits a G3 type star with an orbital period of 4.32 days and has an equilibrium temperature of 1340\,K \citep{Southworth2012}. With a density of only 0.165\,$\mathrm{\rho_{Jup}}$ \citep{Ciceri2013}, WASP-21b is highly inflated, and is one of the lowest density exoplanets discovered. This puffy, extended atmosphere makes WASP-21b an excellent candidate for transmission spectroscopy, with the absorption signal from one atmospheric scale height corresponding to a transit depth of 246\,ppm. 

This paper is organised as follows. In Section \ref{section:obs} we present our observations. In Sections \ref{section:d_red} and \ref{section:fitting} we detail the data reduction and light curve fitting. In Section \ref{section:photometry} we outline photometric monitoring to assess stellar activity. In Section \ref{section:results} the results of the fitting and our combined transmission spectrum are presented. In Section \ref{section:retrieval} we present our atmospheric retrieval analysis. Finally, discussions and conclusions are presented in Sections \ref{section:discussion} and \ref{section:conclusion}.

\section{Observations} \label{section:obs}

Three transits of WASP-21b were observed using the low-resolution grism spectrograph ACAM \citep{Benn2008} on the 4.2\,m William Herschel Telescope (WHT) in La Palma on September 9 and 22 2017 and September 20 2018. The resolution of ACAM for a 1\,arcsec slit at 6000\,\AA\, is $R \approx 450$. However, as detailed below, we used a much wider slit and our resolution is therefore much lower than the quoted value. The nightly resolutions obtained at the \ion{Na}{i} feature are detailed in Section \ref{section:d_red}.

Due to its wide wavelength range ($\sim$3500--9200\,\AA), wide field of view and wide slits, ACAM is well suited to transmission spectroscopy, and is the same instrument previously used for LRG-BEASTS by \citet{Kirk2017, Kirk2018, Kirk2019} and \citet{Louden2017}. The wide, 8\,arcmin field of view and the 40\,arcsec wide, 7.6\,arcmin long slit available on ACAM allow for a greater range of comparison stars to be used for differential photometry, while eliminating the potential for differential slit losses between the target and comparison. The comparison star used across all three nights has co-ordinates 23h09m48.23s +18d22m56.33s, and lies 2.52\,arcmin from WASP-21 at a $V$ magnitude of 11.8 with $(B-V)=0.54$. WASP-21 has a $V$ magnitude of 11.6 and $(B-V)=0.53$, and therefore benefits from having a similar brightness ($\Delta V =0.2$) and colour ($\Delta(B-V)=0.01$) to the comparison.


Both sky and lamp flat-field images were obtained so that the data could be reduced using either type of flat-fielding, allowing the signal-to-noise of the resulting white-light curves to be compared. For Night 1, we obtained 125 biases, 57 sky flats, 123 lamp flats and 155 science spectra all with an exposure time of 200s, except for the first 11 frames, which used an exposure time of 90s. The science observations lasted 8.9 hours and covered an airmass of $1.88\rightarrow1.02\rightarrow2.09$, with the moon $46^{\circ}$ from the target at 84\% illumination. For Night 2, we obtained 222 biases, 108 sky flats, 160 lamp flats and 151 science spectra all with an exposure time of 200s, except for the last 10 frames, which used an exposure time of 300s. The science observations lasted 8.9 hours and covered an airmass of $1.70\rightarrow1.02\rightarrow2.73$, however as some frames were later removed due to cloud coverage, the frames analysed finish at an airmass of 1.07. The moon was $134^{\circ}$ from the target at 8\% illumination. For Night 3, we obtained 153 biases, 124 sky flats, 124 lamp flats and 167 science spectra all with an exposure time of 175s, except for the first 24 frames, which used an exposure time of 200s. The science observations lasted 8.6 hours and covered an airmass of $1.70\rightarrow1.02\rightarrow2.37$, however frames analysed were limited to those below 2.10. The moon was $48^{\circ}$ from the target at 85\% illumination. In all cases, the readout time was 9s. 

Each set of biases was median-combined into a master bias. Each set of flats was also median-combined after removing the master bias, with a running median used in the dispersion direction to ensure sky or lamp lines were removed from each pixel column. The running median was performed over 5 pixels to remove features from the source without removing the pixel-to-pixel sensitivity variations. 


\section{Data Reduction} \label{section:d_red}

To reduce the data, we used our own custom built Python pipeline previously introduced in \citet{Kirk2017,Kirk2018,Kirk2019}. 

With the master biases and flat-fields created as described in Section \ref{section:obs}, the spectral traces of both WASP-21 and the comparison were extracted. To do this, apertures were placed over the traces, with a polynomial fitted to selected background regions either side of each trace to remove the sky background. The positions of the traces were obtained by fitting Gaussians to the trace in the spatial direction, and then fitting a fourth order polynomial to the means of the Gaussians in the dispersion direction. 

We experimented with using no flat-fielding as well as lamp and sky flat-fields, and a variety of extraction aperture widths, background regions and background polynomial orders. The choice of flat-fielding used did not result in significant differences in the RMS of residuals, with differences between reductions typically of the order of 10\,ppm. We found that a combination of no flat-fielding and a second order sky background polynomial produced the lowest RMS in the residuals of the resulting white-light curves. An aperture of 30 pixels was used over each trace for all three transits, with 100 pixel wide background regions used either side of the trace for the first transit, and 40 pixels for the second and third. These combinations were found to result in the lowest RMS of residuals. In all cases, the background regions were offset by 20 pixels from the aperture. The pixel scale of ACAM is 0.253 arsec pixel$^{-1}$. Once the background polynomial had been subtracted, the spectral counts within the apertures were summed, with errors calculated from the read and photon noises.


Diagnostics from the extraction process for each night are shown in Figures \ref{fig:n1_ancillary}, \ref{fig:n2_ancillary} and \ref{fig:n3_ancillary} in the Appendix. As shown in Figures \ref{fig:n2_ancillary} and \ref{fig:n3_ancillary}, cloud coverage during the second and third nights led to the removal of frames with a significant drop in flux from the data set. This resulted in 153 spectra for the first, 68 spectra for the second and 135 spectra for the third night (99\%, 45\%, and 81\% of the respective original totals). Additionally, in the third night's data, we noticed a brightening of the sky during the transit (Fig. \ref{fig:n3_ancillary}, Panels 7 and 8). We believe that this increase in the background is related to moonlight scattered off the telescope structure, as the moon was at 85\% illumination $48^{\circ}$ from the target. However, given the consistency in this night's transmission spectrum with the other two nights (Fig. \ref{fig:trans_spec}, Panel 1), we believe that this scattered light is not impacting our results.

Following the spectral extraction, any cosmic rays were removed from the spectra, with a running median used to identify outliers in the spectra. All of the spectra from one transit were then aligned in pixel space to allow for differential spectrophotometry. Absorption features in the spectra were cross-correlated with a reference spectrum to find the pixel shifts, then the spectra were resampled onto the same wavelength grid as the reference spectrum. The pixel positions of nine spectral features were then calibrated to model and telluric spectral lines to obtain a wavelength solution for the observed spectra.

The spectra from each night were used to obtain the FWHM and resolution at 5894\,\AA\, by fitting a Gaussian to the \ion{Na}{i} absorption feature. For night 1, the FWHM was 30\,\AA, achieving a resolution of 197. For night 2, the FWHM was 27\,\AA, achieving a resolution of 222. For night 3, the FWHM was 16\,\AA, achieving a resolution of 378.


\begin{figure}
    \centering
    \includegraphics[width=\linewidth]{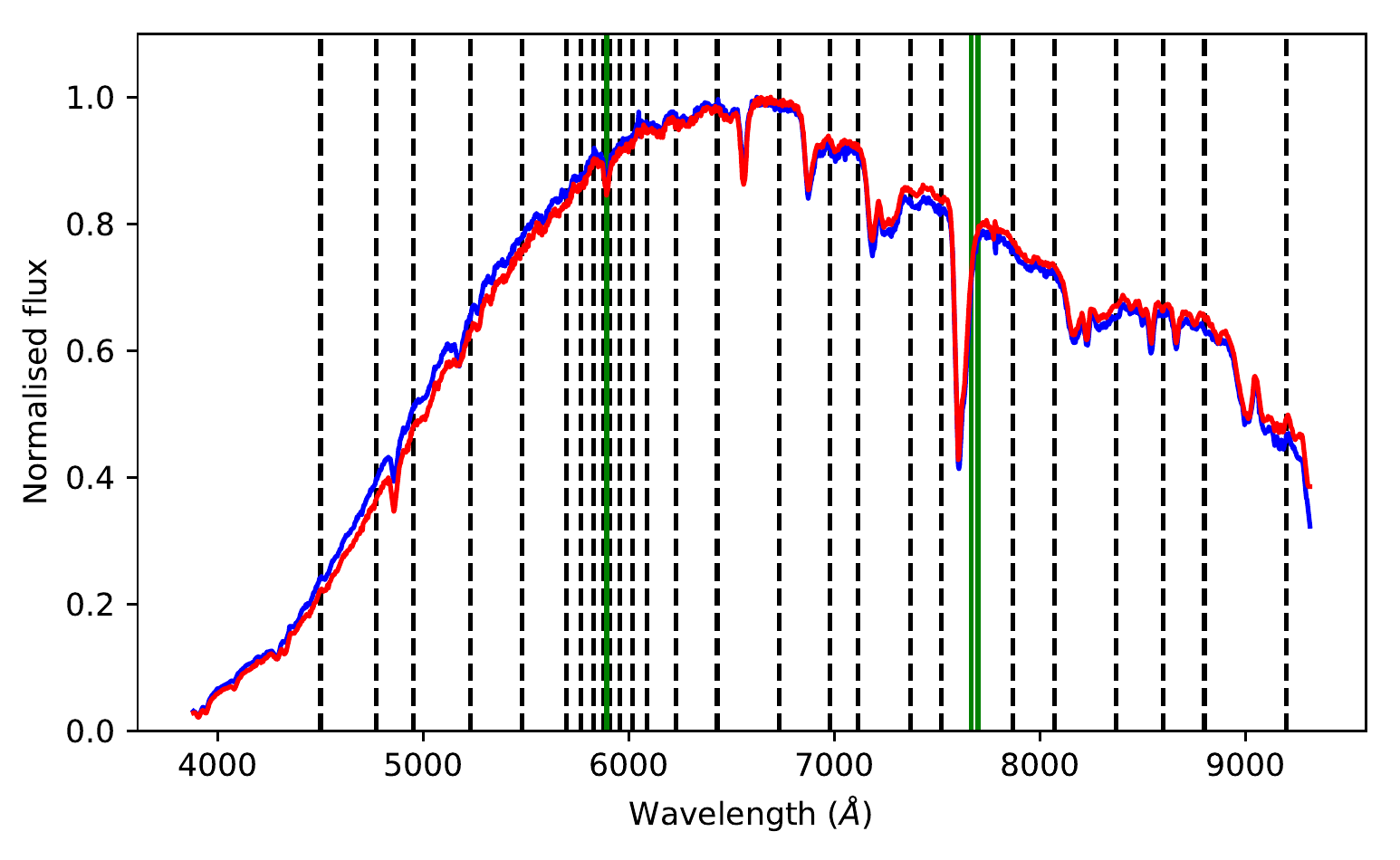}
    \caption{Spectra of WASP-21 (blue line) and the comparison (red line). The black dashed lines show the wavelength bins corresponding to each spectroscopic light curve. The green lines show the location of the \ion{Na}{i} and \ion{K}{i} absorption features.}
    \label{fig:bins}
\end{figure}

As shown in Figure \ref{fig:bins}, the spectra were binned into 25 bins across the range of 4495-9200\,\AA\ so that we could determine the wavelength-dependent transit depths. The locations and widths of the bins were chosen to ensure that the bin edges did not fall onto absorption feature lines, and that the resulting spectroscopic light curves had minimal levels of noise. To assist in the search for sodium and potassium absorption, a narrow 30\,\AA\, bin was also centred on the \ion{Na}{I} doublet (5893\,\AA), and on the right hand edge of the \ion{K}{I} doublet (7699\,\AA) due to the proximity of the telluric O$_{2}$ A band feature at $\sim$7620\,\AA. To look for the wings of the \ion{Na}{i} and \ion{K}{i} absorption features, a series of 18 bins all centred on and increasing incrementally in width from 30-200\,\AA\, were also generated at these locations. A minimum bin size of 30\,\AA\, was used due to the resolution of ACAM, and covers both doublet lines of \ion{Na}{i}. Increasing the bin size beyond this should show a gradual decline in transit depth measured if absorption is detected.

At this point, spectroscopic light curves were made by summing the flux in each wavelength bin, and a white-light curve corresponding to the full wavelength range was made, with the light curves of WASP-21 being divided by the comparison's light curves to correct for telluric absorption. For each transit, a total of 25 spectroscopic light curves were made over the full wavelength range, including light curves corresponding to the additional bins for the \ion{Na}{i} absorption feature.

\section{Light Curve Fitting} \label{section:fitting}

With the white-light and spectroscopic light curves generated following Section \ref{section:d_red}, models were fitted to obtain the parameters of interest, in particular the planet-to-star radius ratio $R_{p} / R_{*}$. 

In order to fit the light curves, analytic transit light curves from \cite{Mandel2002} were fitted using the \verb|Batman| Python package \citep{Kreidberg2015}. A Gaussian process regression (GP) was used to model the systematic noise in the data using the \verb|george| Python package \citep{Ambikasaran2014}. A GP is a non-parametric regression technique which can be used to model correlated noise in data through the use of kernels and hyperparameters to characterise covariance between data points. GPs are becoming increasingly common within exoplanet research, and have proven to be particularly useful in transmission spectroscopy \citep[e.g.][]{Gibson2011,Gibson2012,Evans2015,Evans2017,Louden2017,Kirk2017,Kirk2018,Kirk2019}

For the white-light curve, the time of mid-transit ($T_{c}$), inclination ($i$), ratio of the semi-major axis to the stellar radius ($a/R_{*}$), planet-to-star radius ratio ($R_{p}/R_{*}$) and the linear limb-darkening coefficient ($u_{1}$) were fitted, with wide, uniform priors used to prevent unphysical values. For all light curve fits, the eccentricity was held fixed at zero and the period ($P=4.3225126$ days) was also held fixed to the value from \cite{Seeliger2015}. The quadratic limb-darkening coefficient ($u_{2}$) was held fixed to a theoretical value calculated using the Limb-Darkening Toolkit (\verb|LDTk|) Python package \citep{Parviainen2015}. \verb|LDTk| uses \textsc{phoenix} stellar atmosphere models \citep{Husser2013} to generate limb-darkening coefficients and their uncertainties, given user-defined stellar parameters and uncertainties. The stellar parameters and uncertainties for WASP-21 were taken from \citet{Ciceri2013}.

A maximum of six squared exponential GP kernels with one of airmass, FWHM, mean $x$-pixel position of the stars on the CCD, mean $y$-pixel position of the stars on the CCD, mean sky background and time as an input variable were used. Multiple combinations of kernels were experimented with in order to find the combination of the fewest kernels necessary to fit the noise in the data. Following the methods of \cite{Evans2017,Evans2018}, each of the GP input variables were standardised by subtracting the mean and dividing by the standard deviation, giving each input a mean of zero and a standard deviation of unity. This helps the GP determine the most important inputs for describing the noise characteristics. As the $x$ positions and sky background were obtained as functions of the dispersion direction during the extraction process (Section \ref{section:d_red}), both were binned in wavelength using the same binning scheme as the spectroscopic light curves. In the case of the white-light curve, the $x$ positions and sky background were therefore integrated over the entire wavelength range. Each kernel was defined by a unique length scale ($\tau$), but all shared a common amplitude ($a$). A white noise kernel defined by the variance ($\sigma^{2}$) was also included. Similar to \citet{Gibson2017}, we placed truncated uniform priors in log space on the GP hyperparameters. The GP amplitude was bounded by 0.5 and 2$\times$ the variance of the out-of-transit data, and the length scales were bounded by the minimum spacing between the standardised data points and 5$\times$ the full span of each standardised variable. We also tested other sets of priors, and found that the results were consistent. 

Fitting was performed using a Markov chain Monte Carlo (MCMC) method using the \verb|emcee| Python package \citep{Foreman-Mackey2013} with $12\times N$ walkers each performing 10,000 steps, where $N$ is the number of parameters. For Night 1, the best white-light curve fit resulted from using a time kernel only, and for Nights 2 and 3, time and FWHM kernels, resulting in $N=8$ for Night 1, and $N=9$ for Nights 2 and 3. The addition of the FWHM kernel for Nights 2 and 3 follows the cloud coverage observed (Figs. \ref{fig:n2_ancillary}, \ref{fig:n3_ancillary}) as the FWHM is also affected by the presence of cloud. The walkers were initialised with a small scatter around the starting values for each parameter, and the first 5000 steps of each walker were discarded as burn-in. Following the \verb|george| documentation, a second chain was then run with 10,000 steps, again discarding the first 5000 as burn-in with the walkers initially scattered around the median values from the first chain. Before executing the MCMC, a running median was used to clip any outlying points from the transit light curves which deviated by >4$\sigma$ from the median, removing at most 1-2 points per light curve. The starting locations for the GP hyperparameters were obtained by optimising the hyperparameters to the out-of-transit data, while \citet{Ciceri2013} provided starting locations for the transit light curve parameters, and \verb|LDTk| for $u_{1}$.

The process was then repeated for each of the spectroscopic light curves, but with $a/R_{*}$, $T_{c}$ and inclination fixed to the results of the best fitting white-light curve fit in order to minimise the uncertainties in $R_{p}/R_{*}$. This resulted in a total of 5 parameters for Night 1 and 6 for Nights 2 and 3, each again with 12 corresponding walkers. Iterations of different kernel combinations were again compared to obtain the best fits to the noise. A common mode correction obtained from the systematics model of the white-light curve was applied to the spectroscopic light curves before fitting to remove systematics that are common to the spectroscopic light curves, helping to increase the precision in $R_{p}/R_{*}$. 

MCMCs were run for each spectroscopic light curve using the same method as the white-light curves. For the first and second night, the same kernels were determined as the best fitting for the spectroscopic light curves as their respective white-light curves, however for the third night, a combination of sky background and FWHM was found to be optimal. While including a time kernel led to a better agreement in $R_{p}/R_{*}$ for the white-light curve fits, doing so for the spectroscopic light curves led to a poor fit to noise, and a transmission spectrum in poor agreement with the other nights. Given the scattered light in the sky background as mentioned in Section \ref{section:d_red}, we were motivated to use a sky background kernel to model the noise, and we note that using the system parameters obtained from a sky and FWHM white-light fit results in a consistent transmission spectrum, but offset in $R_{p}/R_{*}$. Further analysis of binned light curves for each of the nights, for example those corresponding with the incrementally increasing bins around the absorption features, followed the same process and kernel combinations as the respective spectroscopic light curve fits.

\section{Photometric Monitoring for Stellar Activity} \label{section:photometry}

\begin{figure}
    \centering
    \includegraphics[width=\linewidth]{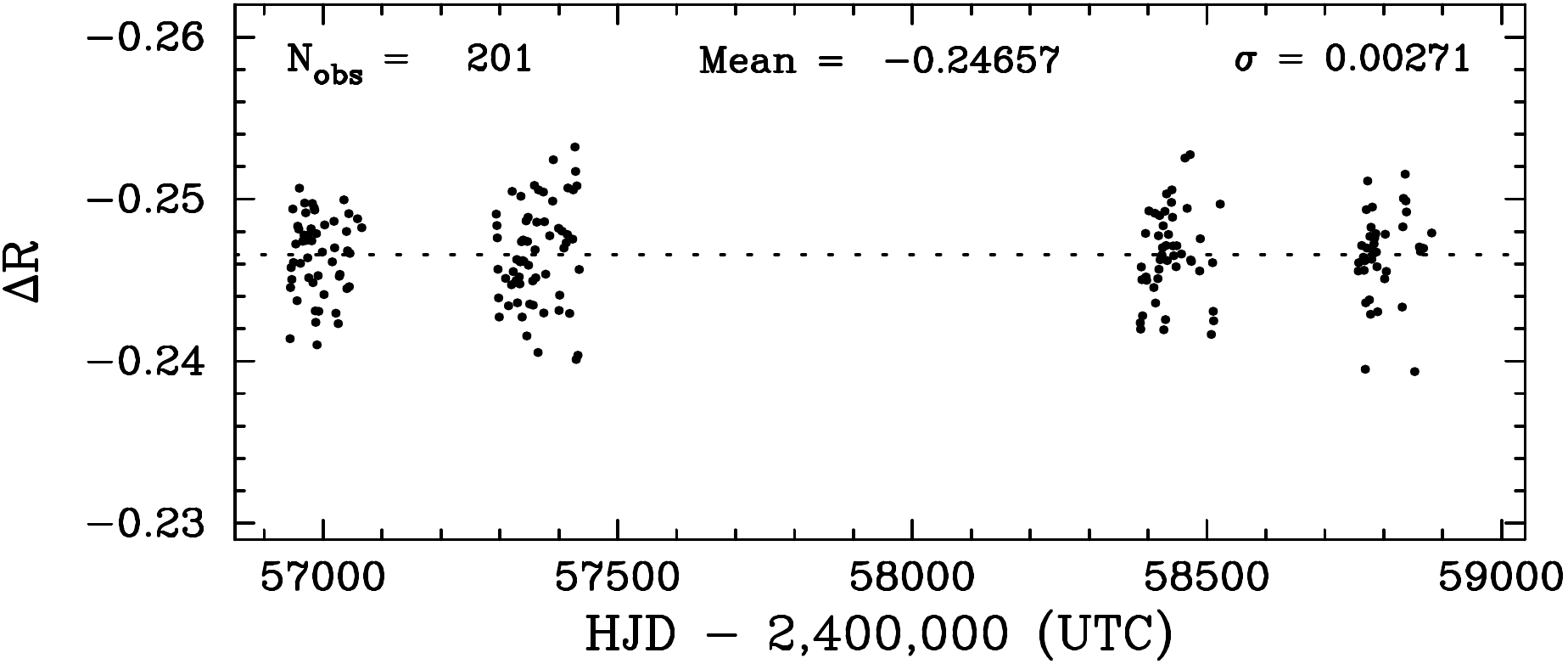}
    \caption{Photometric data of WASP-21 obtained with the C14-AIT at Fairborn Observatory. Individual observing seasons have been normalized to have the same seasonal mean (given as the dotted line). The standard deviation of the data set about its mean is only 0.00271 mag, consistent with the precision of a single observation from the AIT. We conclude that WASP-21 is an inactive star with little photometric variability.}
    \label{fig:periodogram}
\end{figure}

To monitor and characterise the stellar activity of WASP-21, we acquired four years of photometry with the Tennessee State University Celestron 14-inch Automated Imaging Telescope (C14-AIT) at Fairborn Observatory in Arizona \citep{Henry1999,Eaton2003}. A total of 201 observations of WASP-21 were collected between the 2014-15 and 2019-20 observing seasons, and are summarised in Table \ref{tab:photometry} and plotted in Figure \ref{fig:periodogram}. Due to the failure and 
replacement of the AIT's CCD camera between the second and third seasons, we normalised the individual seasons to have the same mean brightness to remove a small systematic shift in the star's brightness. Further details of our 
data acquisition, reduction, and analysis procedures can be found in \cite{Sing2015}, who describe similar observations of another planetary host star, WASP-31. 

\begin{table}
\centering
    \caption{Summary of C14-AIT photometric observations of WASP-21}
    \label{tab:photometry}
\begin{tabular}{ccccc}
\hline
Observing & & Date Range & Sigma & Seasonal Mean \\
Season & $N_\mathrm{obs}$ & (HJD-2400000) & (mag) &  (mag) \\ \hline
2014-15 & 54 & 56943-57065 & 0.00235 & $-0.24657\pm0.00032$ \\
2015-16 & 61 & 57293-57434 & 0.00308 & $-0.24379\pm0.00039$ \\
2018-19 & 47 & 58386-58522 & 0.00273 & $-0.25445\pm0.00040$ \\
2019-20 & 39 & 58756-58881 & 0.00265 & $-0.25013\pm0.00043$ \\
\hline
\end{tabular}
\end{table}

Figure \ref{fig:periodogram} gives the standard deviation of the individual observations from the normalised mean (shown as a dotted line) to be 0.00271 mag, comparable to the AIT's measurement precision as determined from pairs of constant stars. Periodogram analyses of the four individual observing seasons as well as the whole data set find no significant periodicities. We conclude that WASP-21 is constant to the limit of our precision and so is likely an inactive star with little photometric variability. This result reflects the previous 
findings of \cite{Ciceri2013}, who also report that WASP-21 has low activity levels. However, low photometric modulation does not necessarily imply low activity \citep{Rackham2018}, therefore we consider the effects of stellar activity on our transmission spectrum in Section \ref{section:activity}.

\section{Results} \label{section:results}

\begin{table*}
    \centering
    \caption{Results from the fitting of the white light curves (Fig. \ref{fig:wl_curves}) using a Gaussian process for each of the three nights. Note that the quadratic limb-darkening coefficient ($u_{2}$) and the period ($P$) were held fixed. We also include the weighted mean of the three nights, and results from \citet{Seeliger2015}. }
    \label{tab:wl_bestparams}
\begin{tabular}{cccccc}
\hline
Parameter & Night 1 & Night 2 & Night 3 & Weighted Mean & \cite{Seeliger2015} \\ \hline
$P$ (days) & 4.3225126 & - & - & - & $4.3225126\pm0.0000022$\\
$a/R_{*}$ & 9.36$^{+0.29}_{-0.26}$ & 9.46$^{+0.25}_{-0.26}$ & 8.92$^{+0.22}_{-0.22}$ & $9.41\pm0.19$ & $9.62\pm0.17$ \\ 
$i$ ($^\circ$) & 86.86$^{+0.41}_{-0.35}$ & 86.80$^{+0.31}_{-0.32}$ & 86.23$^{+0.29}_{-0.29}$ & $86.82\pm0.24$ & $87.12\pm0.24$ \\
$T_{c}$ (BJD) & 2458006.53493$^{+0.00023}_{-0.00023}$ & 2458019.50152$^{+0.00030}_{-0.00028}$ & 2458382.59276$^{+0.00031}_{-0.00029}$ & - & $2454743.04217\pm0.00065$ \\
$R_{p}/R_{*}$ & 0.10319$^{+ 0.00200}_{-0.00190}$ & 0.10255$^{+0.00186}_{-0.00175}$ & 0.10020$^{+0.00150}_{-0.00152}$ & $0.10280\pm0.00132$ & $0.1030\pm0.0008$ \\
$u_{1}$ & 0.48$^{+0.06}_{-0.07}$ & 0.41$^{+0.05}_{-0.05}$ & 0.43$^{+0.07}_{-0.08}$ & $0.44\pm0.03$ & - \\
$u_{2}$ & $0.14$ & $0.14$ & $0.14$ & - & - \\ \hline

\end{tabular}
\end{table*}

\begin{table*}
\centering
\caption{Results of the wavelength-binned fits (Figures \ref{fig:n1_sl_curves}, \ref{fig:n2_sl_curves}, and \ref{fig:n3_sl_curves}) and the combined values of $R_{P} / R_{*}$ along with their relative uncertainties. Note that the $R_{P} / R_{*}$ values for Nights 2 and 3 have been median offset to that of Night 1.}
\label{tab:3ncombined}
\begin{tabular}{cccccc}
\hline
Bin & Bin & Night 1 & Night 2 & Night 3 & \textbf{Combined} \\
Centre (\AA) & Width (\AA) & $R_{P} / R_{*}$ & $R_{P} / R_{*}$ & $R_{P} / R_{*}$ & $\mathbf{R_{P} / R_{*}}$  \\ \hline
4635.0 & 270.0 & $0.10518^{+0.00184}_{-0.00244}$ & $0.10745^{+0.00220}_{-0.00242}$ & $0.10294^{+0.00106}_{-0.00119}$ & $\mathbf{0.10664 \pm 0.00150}$ \\
4860.0 & 180.0 & $0.10684^{+0.00087}_{-0.00089}$ & $0.10322^{+0.00203}_{-0.00244}$ & $0.10282^{+0.00150}_{-0.00124}$ & $\mathbf{0.10645 \pm 0.00082}$ \\
5090.0 & 280.0 & $0.10596^{+0.00091}_{-0.00108}$ & $0.10424^{+0.00107}_{-0.00117}$ & $0.10401^{+0.00071}_{-0.00068}$ & $\mathbf{0.10548 \pm 0.00074}$ \\
5355.0 & 250.0 & $0.10573^{+0.00090}_{-0.00107}$ & $0.10290^{+0.00121}_{-0.00128}$ & $0.10309^{+0.00061}_{-0.00066}$ & $\mathbf{0.10489 \pm 0.00077}$ \\
5589.0 & 218.0 & $0.10133^{+0.00128}_{-0.00134}$ & $0.10345^{+0.00105}_{-0.00108}$ & $0.10284^{+0.00049}_{-0.00049}$ & $\mathbf{0.10292 \pm 0.00083}$ \\
5733.0 & 70.0  & $0.10208^{+0.00208}_{-0.00209}$ & $0.099656^{+0.00139}_{-0.00157}$ & $0.10466^{+0.00071}_{-0.00069}$ & $\mathbf{0.10085 \pm 0.00121}$ \\
5798.0 & 60.0  & $0.10399^{+0.00209}_{-0.00213}$ & $0.10357^{+0.00147}_{-0.00125}$ & $0.10367^{+0.00074}_{-0.00076}$ & $\mathbf{0.10406 \pm 0.00122}$ \\
5853.0 & 50.0  & $0.10202^{+0.00225}_{-0.00218}$ & $0.10487^{+0.00226}_{-0.00222}$ & $0.10286^{+0.00087}_{-0.00083}$ & $\mathbf{0.10366 \pm 0.00157}$ \\
5893.0 & 30.0  & $0.10784^{+0.00161}_{-0.00235}$ & $0.10979^{+0.00170}_{-0.00170}$ & $0.10543^{+0.00106}_{-0.00106}$ & $\mathbf{0.10939 \pm 0.00129}$ \\
5933.0 & 50.0  & $0.10468^{+0.00191}_{-0.00224}$ & $0.10359^{+0.00304}_{-0.00277}$ & $0.10400^{+0.00072}_{-0.00073}$ & $\mathbf{0.10453 \pm 0.00169}$ \\
5988.0 & 60.0  & $0.10198^{+0.00234}_{-0.00225}$ & $0.10081^{+0.00181}_{-0.00220}$ & $0.10351^{+0.00063}_{-0.00066}$ & $\mathbf{0.10168 \pm 0.00151}$ \\
6053.0 & 70.0  & $0.10224^{+0.00131}_{-0.00157}$ & $0.10326^{+0.00228}_{-0.00210}$ & $0.10259^{+0.00072}_{-0.00073}$ & $\mathbf{0.10275 \pm 0.00120}$ \\
6159.0 & 142.0 & $0.10376^{+0.00162}_{-0.00151}$ & $0.10408^{+0.00099}_{-0.00100}$ & $0.10278^{+0.00048}_{-0.00048}$ & $\mathbf{0.10434 \pm 0.00084}$ \\
6330.0 & 200.0 & $0.10327^{+0.00081}_{-0.00089}$ & $0.10246^{+0.00087}_{-0.00091}$ & $0.10199^{+0.00046}_{-0.00045}$ & $\mathbf{0.10315 \pm 0.00061}$ \\
6580.0 & 300.0 & $0.10340^{+0.00071}_{-0.00093}$ & $0.10173^{+0.00086}_{-0.00086}$ & $0.10209^{+0.00039}_{-0.00042}$ & $\mathbf{0.10289 \pm 0.00059}$ \\
6855.0 & 250.0 & $0.10281^{+0.00175}_{-0.00098}$ & $0.10244^{+0.00072}_{-0.00072}$ & $0.10169^{+0.00040}_{-0.00041}$ & $\mathbf{0.10285 \pm 0.00064}$ \\
7047.0 & 135.0 & $0.10153^{+0.00062}_{-0.00064}$ & $0.10282^{+0.00096}_{-0.00095}$ & $0.10316^{+0.00054}_{-0.00055}$ & $\mathbf{0.10208 \pm 0.00052}$ \\
7242.0 & 255.0 & $0.10096^{+0.00108}_{-0.00102}$ & $0.10191^{+0.00081}_{-0.00092}$ & $0.10176^{+0.00052}_{-0.00053}$ & $\mathbf{0.10184 \pm 0.00067}$ \\
7445.0 & 150.0 & $0.10190^{+0.00098}_{-0.00102}$ & $0.10156^{+0.00121}_{-0.00145}$ & $0.10133^{+0.00061}_{-0.00062}$ & $\mathbf{0.10200 \pm 0.00080}$ \\
7695.0 & 350.0 & $0.10282^{+0.00096}_{-0.00089}$ & $0.10138^{+0.00093}_{-0.00111}$ & $0.10079^{+0.00056}_{-0.00057}$ & $\mathbf{0.10241 \pm 0.00069}$ \\
7970.0 & 200.0 & $0.10181^{+0.00057}_{-0.00055}$ & $0.10101^{+0.00114}_{-0.00118}$ & $0.09870^{+0.00061}_{-0.00064}$ & $\mathbf{0.10176 \pm 0.00050}$ \\
8220.0 & 300.0 & $0.10248^{+0.00131}_{-0.00111}$ & $0.10165^{+0.00116}_{-0.00138}$ & $0.10117^{+0.00068}_{-0.00067}$ & $\mathbf{0.10232 \pm 0.00088}$ \\
8485.0 & 230.0 & $0.10277^{+0.00067}_{-0.00064}$ & $0.10173^{+0.00237}_{-0.00160}$ & $0.09938^{+0.00087}_{-0.00090}$ & $\mathbf{0.10268 \pm 0.00062}$ \\
8700.0 & 200.0 & $0.10305^{+0.00133}_{-0.00092}$ & $0.10301^{+0.00122}_{-0.00134}$ & $0.10255^{+0.00080}_{-0.00078}$ & $\mathbf{0.10318 \pm 0.00085}$ \\
9000.0 & 400.0 & $0.10414^{+0.00194}_{-0.00185}$ & $0.10221^{+0.00129}_{-0.00131}$ & $0.10423^{+0.00093}_{-0.00094}$ & $\mathbf{0.10317 \pm 0.00107}$ \\ \hline
\end{tabular}
\end{table*}

\begin{figure}
    \centering
    \includegraphics[width=\linewidth]{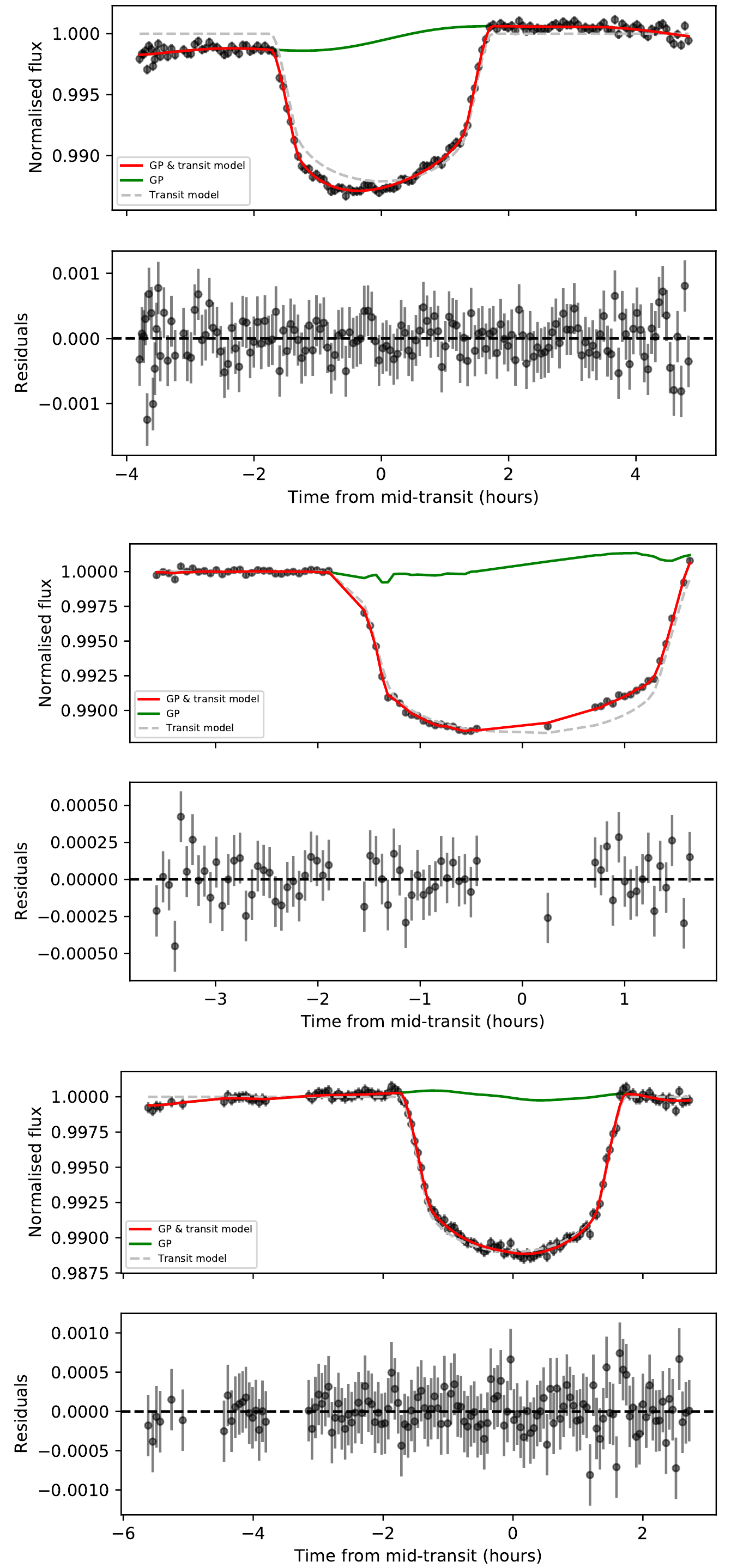}
    \caption{Plots of the fits of the transit white-light curves for Night 1, 2 and 3 respectively. Upper panels: the data points are shown in black, with the best fitting model shown in red. The green and grey lines show the respective contributions from the GP and the analytic transit light curve models. Lower panels: the residuals of the best-fitting model as shown in the corresponding upper panels.}
    \label{fig:wl_curves}
\end{figure}

The fitted white-light curves for each of the three transits are shown in Figure \ref{fig:wl_curves}, with the best fit parameters and values used for $u_{2}$ listed in Table \ref{tab:wl_bestparams}. Table \ref{tab:wl_bestparams} also includes results from \cite{Seeliger2015}, which show a good agreement with our results. 

\begin{figure*}
    \centering
    \includegraphics[width=\linewidth]{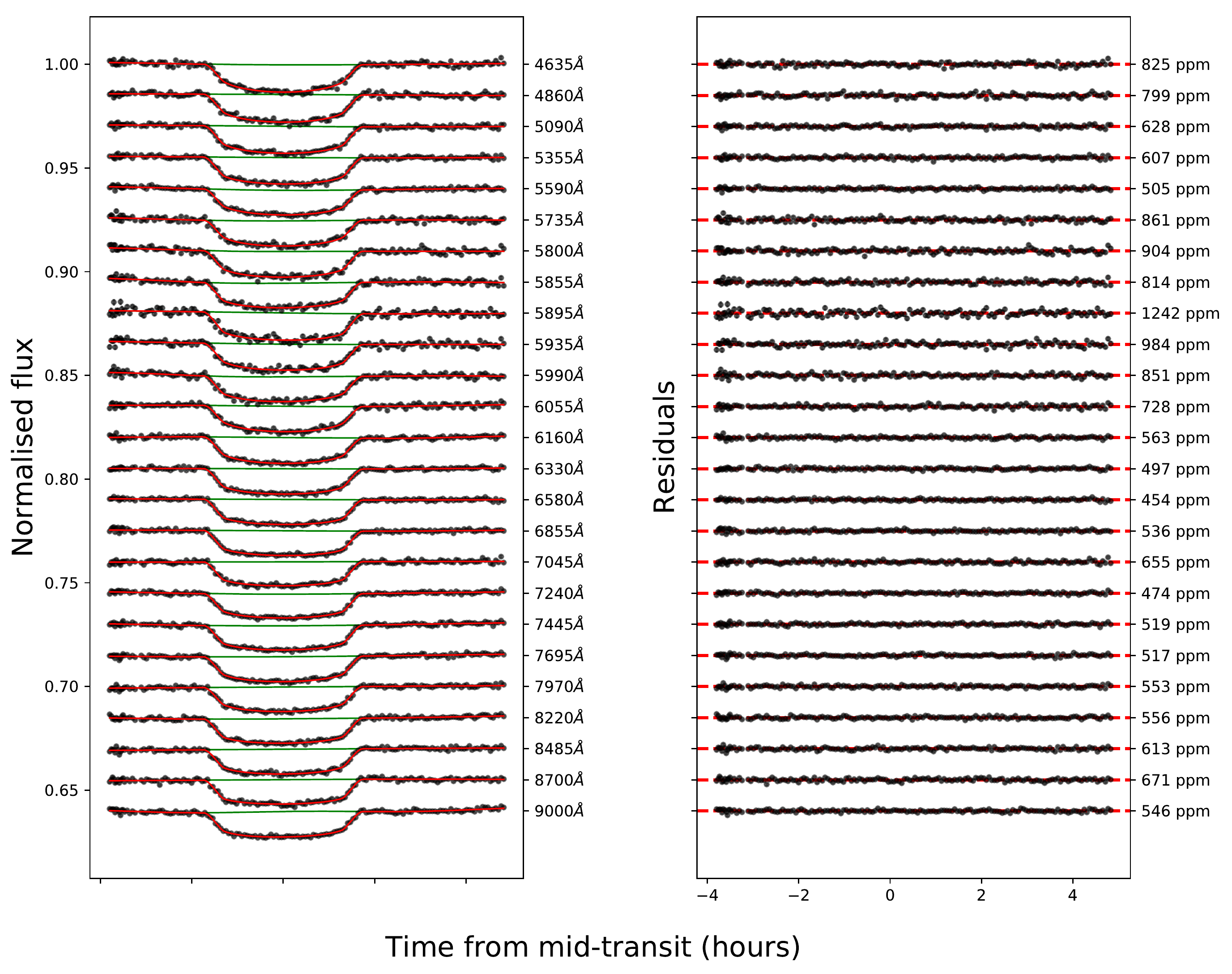}
    \caption{Fits of the spectroscopic light curves for the first transit using an analytic transit light curve and a GP. Left panel: the data points are shown in black, with each light curve offset in $y$ for clarity. The central wavelength of each wavelength bin is shown on the right-hand $y$-axis. Each best fitting model is shown in red. The green lines shows the contributions from the GP models.  Right panel: the residuals of the fits as shown in the left hand panel, with the RMS of residuals given on the right-hand $y$-axis.}
    \label{fig:n1_sl_curves}
\end{figure*}

\begin{figure*}
    \centering
    \includegraphics[width=\linewidth]{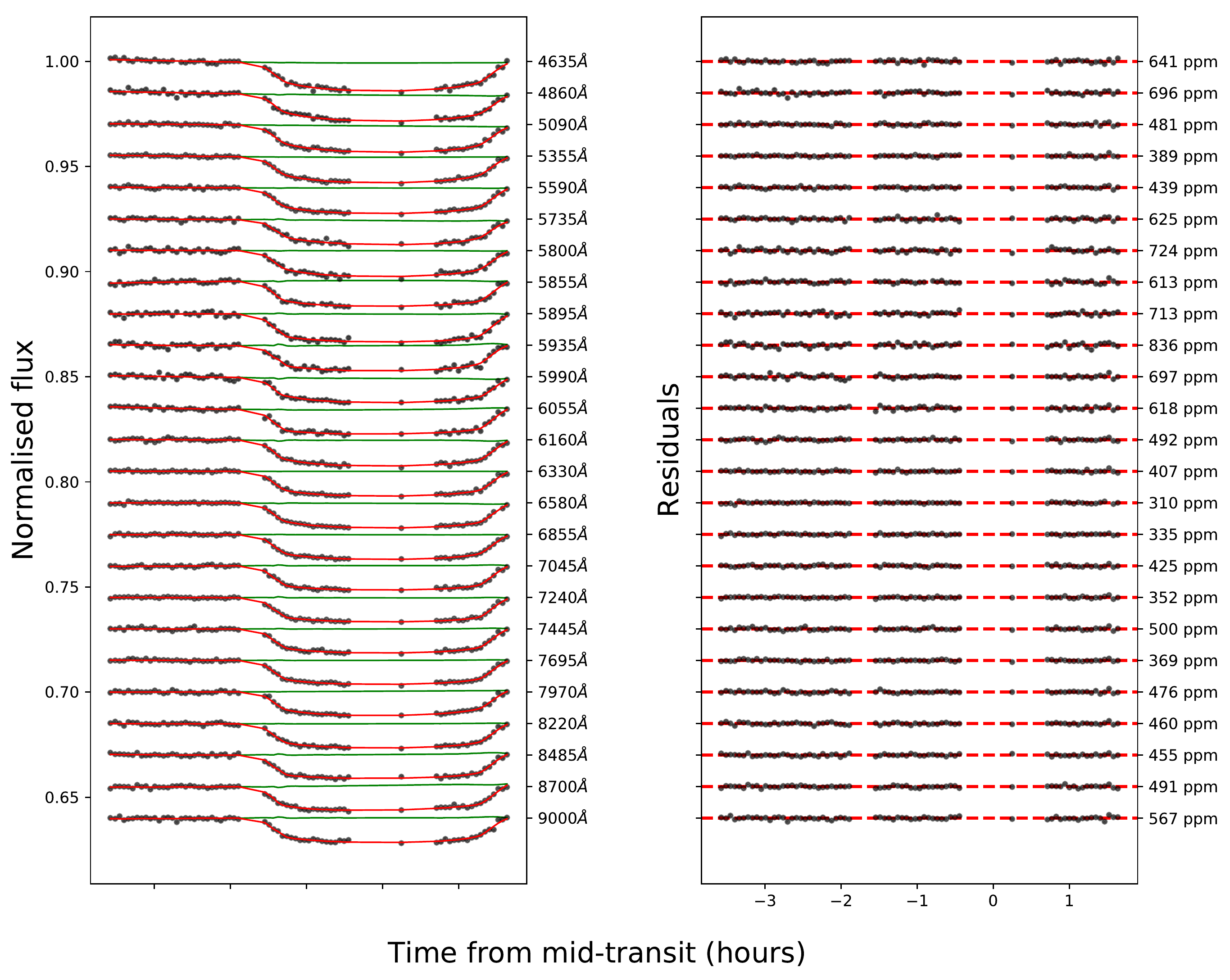}
    \caption{Fits of the spectroscopic light curves for the second transit. For details see Figure \ref{fig:n1_sl_curves}.}
    \label{fig:n2_sl_curves}
\end{figure*}

\begin{figure*}
    \centering
    \includegraphics[width=\linewidth]{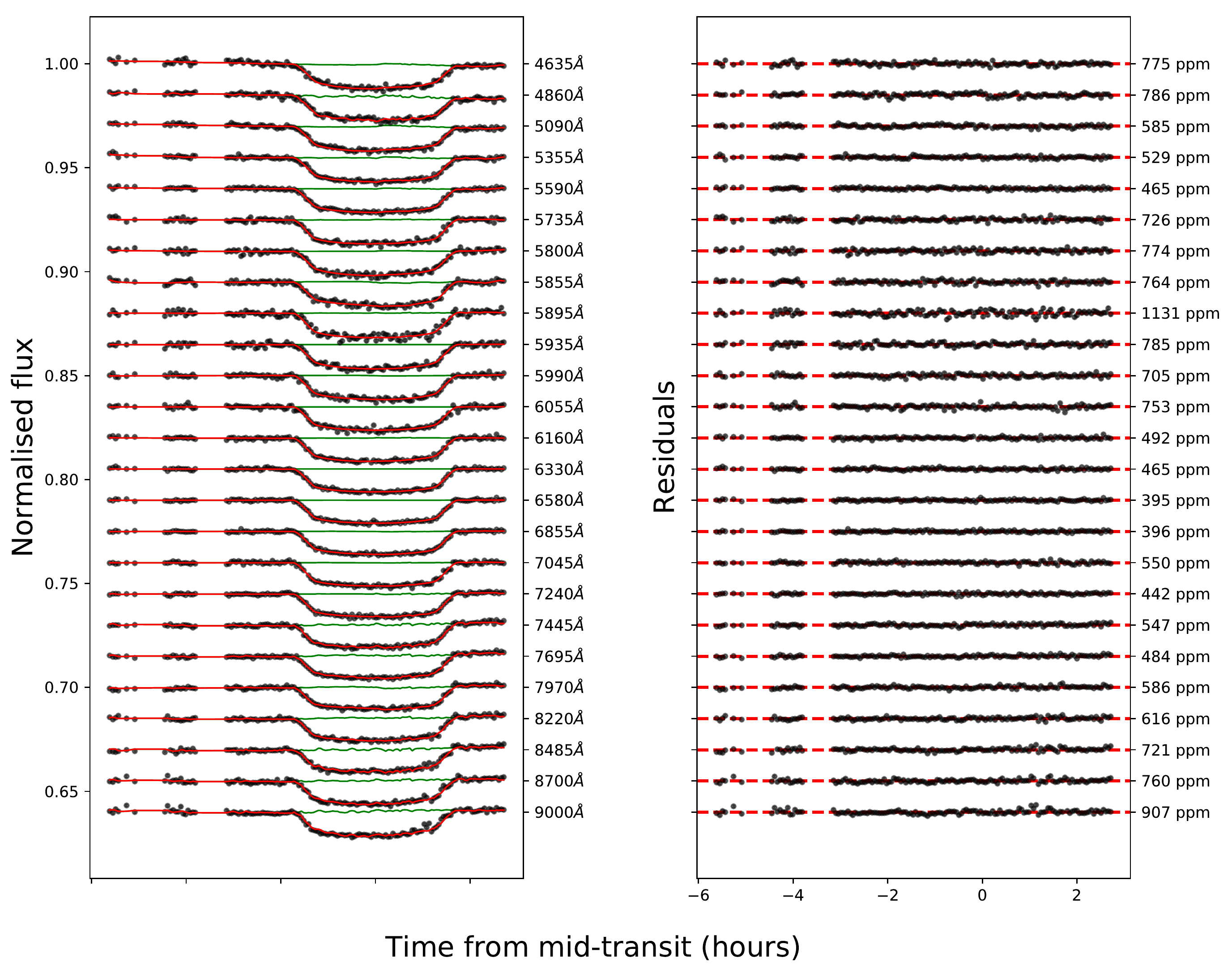}
    \caption{Fits of the spectroscopic light curves for the third transit. For details see Figure \ref{fig:n1_sl_curves}.}
    \label{fig:n3_sl_curves}
\end{figure*}

The fitted spectroscopic light curves are shown in Figure \ref{fig:n1_sl_curves} for the first night, Figure \ref{fig:n2_sl_curves} for the second night and Figure \ref{fig:n3_sl_curves} for the third night. The resulting $R_{P} / R_{*}$ values are given in Table \ref{tab:3ncombined}.

\subsection{Transmission Spectrum}

\begin{figure}
    \centering
    \includegraphics[width=\linewidth]{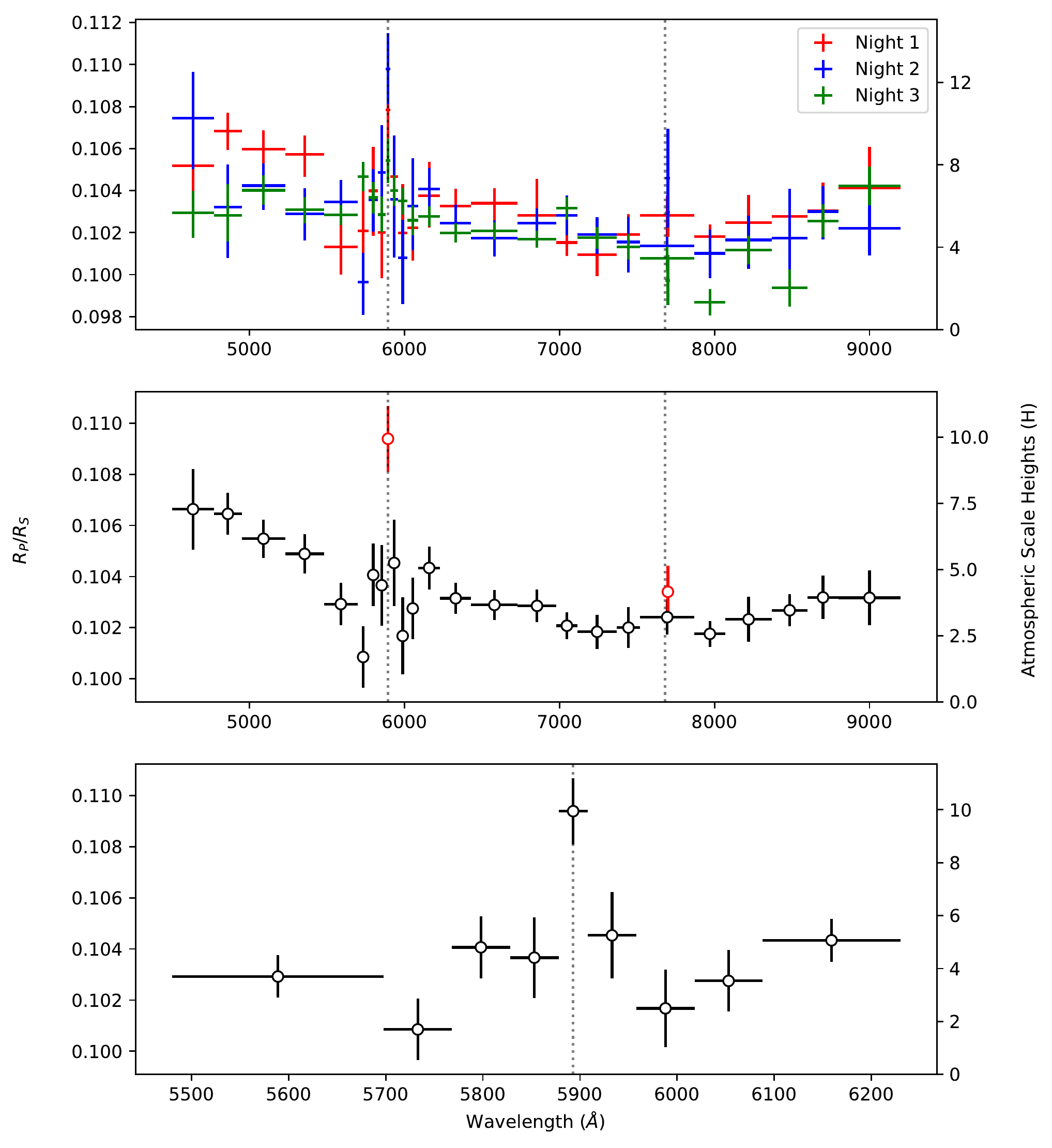}
    \caption{Top panel: the transmission spectra resulting from the spectroscopic light curve fits from Night 1 (red error bars), Night 2 (blue error bars) and Night 3 (green error bars). Middle panel: the combined transmission spectrum resulting from the weighted mean of the three nights. The narrowest 30\,\AA\, bins for the \ion{Na}{i} (5893\,\AA) and \ion{K}{i} (7699\,\AA) absorption features are plotted in red. Bottom panel: the combined transmission spectrum resulting from the weighted mean of the three nights, zoomed to show the \ion{Na}{i} absorption feature at 5893\,\AA. In each panel the dotted grey lines show the centres of the \ion{Na}{i} and \ion{K}{i} doublets at 5893\,\AA\, and 7697\,\AA\, respectively. We note that the potassium bin is centred on the right hand edge of the \ion{K}{i} doublet, due to the telluric oxygen feature which lies just blueward of the \ion{K}{i} doublet, and therefore this bin does not lie directly on the corresponding grey line. The scale height of WASP-21b was calculated using parameters obtained from \citet{Ciceri2013} assuming an H/He atmosphere with a mean molecular mass of $\mu=2.3$. }
    \label{fig:trans_spec}
\end{figure}

After fitting the spectroscopic light curves for each of the three nights, individual transmission spectra were constructed for each night, along with a combined transmission spectrum using the weighted mean of the individual nights (Fig. \ref{fig:trans_spec}). To do so, the values of $R_{P}/R_{*}$ for the second and third night were median offset to that of the first night to account for small differences in the mean $R_{P}/R_{*}$ (Table \ref{tab:wl_bestparams}), as shown in the top panel of Figure \ref{fig:trans_spec}. The combined values of $R_{P}/R_{*}$ are listed in Table \ref{tab:3ncombined}. 

The middle panel of Figure \ref{fig:trans_spec} shows the combined transmission spectrum to have a strong scattering slope and sodium absorption (shown by the red bin at 5893\,\AA), but no detectable potassium absorption in a bin width of 30\,\AA\, (shown by the red bin at 7699\,\AA). Given the low photometric activity observed in WASP-21 (Section \ref{section:photometry}), we believe that this slope is most likely due to aerosols and not stellar activity. However, all possible explanations of this slope are explored in Section \ref{section:hd189}. A closer view of the sodium absorption feature is shown in the lower panel of Figure \ref{fig:trans_spec}. We discuss this in more detail in the following section.

\subsection{Sodium and Potassium} \label{section:sodium}

To further explore the presence of absorption from the sodium and potassium resonance doublets, narrow bins and bins incrementally increasing in width were produced as described in Section \ref{section:d_red}. 

\begin{figure}
    \centering
    \includegraphics[width=\linewidth]{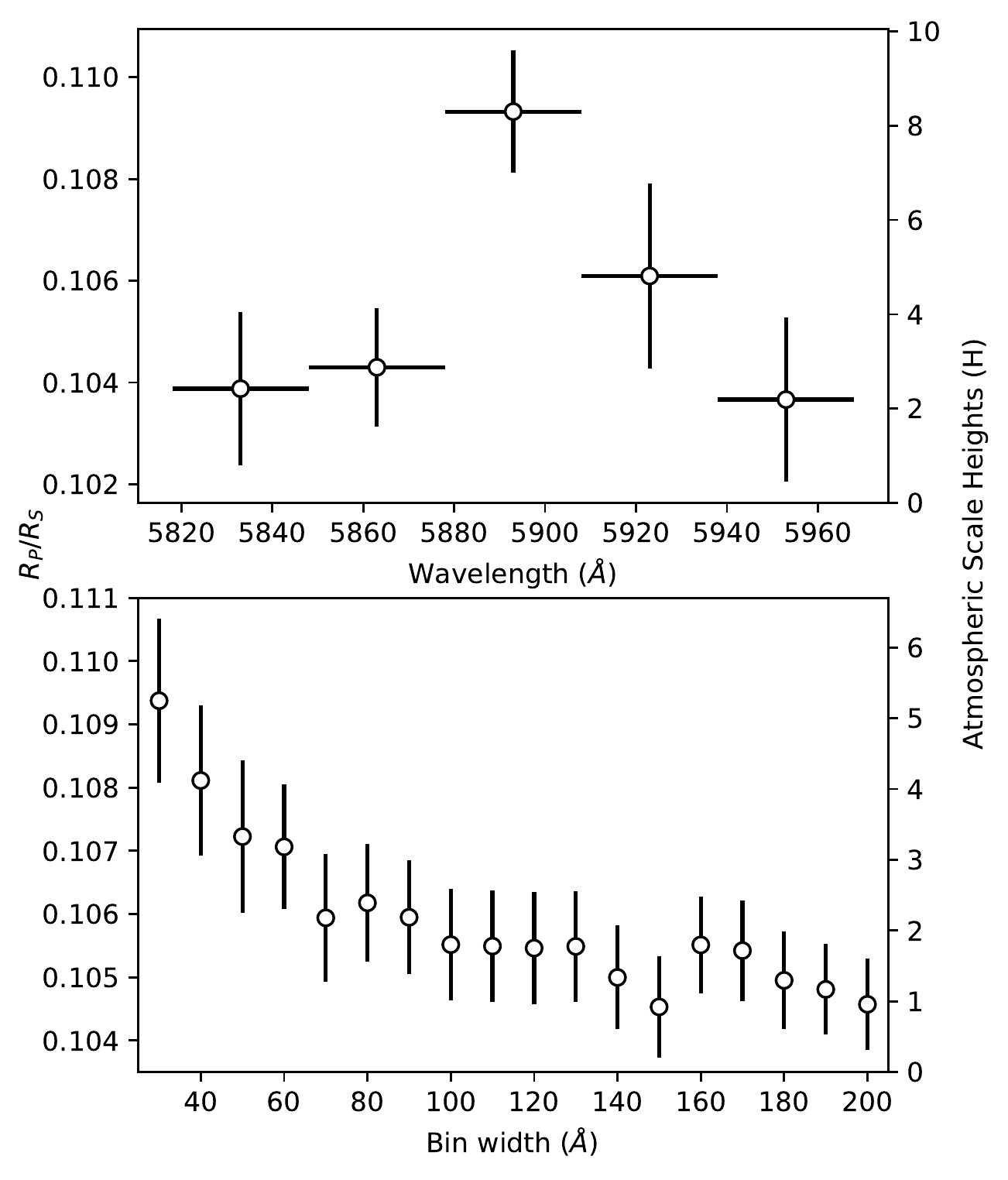}
    \caption{Top panel: Narrow, 30\,\AA\, bins focused on the central wavelength of \ion{Na}{i} at 5893\,\AA. Bottom panel: \ion{Na}{i} absorption feature at 5893\,\AA\, shown in bins of incrementally increasing width from 30-200\,\AA.}
    \label{fig:iib}
\end{figure}

In the case of the \ion{Na}{i} feature, narrow bins focused around the doublet show an increase in the measured transit depth when compared with the continuum, as seen in Figure \ref{fig:trans_spec}. The lower panel of Figure \ref{fig:iib} also shows a detection of the \ion{Na}{i} feature, with a gradual reduction in the measured transit depth from bins of width 30-100\,\AA, after which the signal extends into the continuum, indicating that the absorption feature extends to 100\,\AA. The presence of \ion{Na}{i} absorption at 30\,\AA\, also motivated us to study the feature with an additional series of 30\,\AA\, bins between 5833 and 5953\,\AA\, for each night. The results of the fitting of the corresponding light curves were combined in the same way as the main transmission spectrum and are shown in the upper panel of Figure \ref{fig:iib}. 

To estimate the significance of the \ion{Na}{i} detection, we followed a similar process to that of \cite{Nikolov2016} and \cite{Carter2019}. We fitted a straight line to the slope observed in each of our transmission spectra from wavelengths of 4635-7242\,\AA\,, excluding the 30\,\AA\, bin. By comparing the transit depth measured in the 30\,\AA\, bin to that at the fitted continuum, the significance of the detection of the core of the feature was obtained. Table \ref{tab:na_sig} gives the confidences for each data set, both for the transmission spectra shown in Figure \ref{fig:trans_spec}, and with stellar activity corrections applied, as further detailed in Section \ref{section:activity}. In all cases, the combined transmission spectrum has a significance of detection of the \ion{Na}{i} feature greater than 3$\sigma$, with the uncorrected spectrum at a confidence of 4.03$\sigma$. 

The above procedures were repeated for \ion{K}{i}, with the narrowest bin centred on the right hand edge of the doublet due to the proximity of the oxygen telluric feature. We do not see evidence for \ion{K}{i} absorption, noted in Figure \ref{fig:trans_spec} by the transit depth of the red point at 7699\,\AA\, being consistent with the neighbouring continuum. An additional search was performed for H$\alpha$, however we did not detect any signs of this to a resolution of 30\,\AA.

The implications of the detection of \ion{Na}{i} absorption, and the lack of \ion{K}{I} absorption are discussed further in Section \ref{section:na_dis}.

\begin{table}
\centering
    \caption{Sigma confidence levels of the \ion{Na}{i} absorption feature for the individual and combined data sets with respect to the transmission spectrum baselines. The confidence levels are calculated with and without the application of stellar activity corrections, see Section \ref{section:activity} for further details. In all cases, a significant ($>3\sigma$) \ion{Na}{i} detection is made in the combined transmission spectrum.}
    \label{tab:na_sig}
\begin{tabular}{cccc}
\hline
 & No & Activity & Activity \\
 & Activity & Corrected & Corrected \\
 & Correction & (free slope) & (fixed slope) \\
 & ($\sigma$) & ($\sigma$) & ($\sigma$) \\ \hline
Combined & 4.03 & 3.50 & 3.54 \\
Night 1 & 2.18 & 1.39 & 1.35 \\ 
Night 2 & 3.91 & 2.76 & 2.82 \\ 
Night 3 & 2.61 & 2.00 & 1.83 \\ \hline
\end{tabular}
\end{table}

\section{Atmospheric Retrieval Analysis} \label{section:retrieval}

In this section we present retrieval analyses of the transmission spectra presented in Section \ref{section:results}, with the aim of constraining the scattering slope seen in Figure \ref{fig:trans_spec} and testing the possible effects of stellar activity. We ran retrievals on all the data sets listed in Table \ref{tab:3ncombined} to compare both the individual nights' transmission spectra and the combined spectrum.

Retrievals were performed using the \verb|PLATON| Python package \citep{Zhang2019}, an open-source Python code which assumes equilibrium chemistry. \verb|PLATON| allows the effects of stellar activity to be accounted for, using the temperature of the active regions ($T_\mathrm{active}$) and the covering fraction ($f_\mathrm{active}$), interpolating stellar models and correcting the transit depth for the contribution of the spectrum of the active region. Clouds and hazes are also considered using the logarithm of the cloud-top pressure ($\log P_\mathrm{cloud}$), the gradient of the scattering slope ($\alpha$) and the logarithm of a factor that multiplies the scattering slope, adjusting it in transit depth ($\log s$). 

These parameters, along with the planet's radius at a pressure of 1 bar ($R_\mathrm{P}$), the mass of the planet ($M_\mathrm{P}$), the limb temperature ($T_\mathrm{limb}$) and the logarithm of the atmospheric metallicity relative to the solar value ($\log Z$), were retrieved in a variety of combinations in order to assess the impact of the scattering slope and any stellar activity. The key combinations of parameters used for the models and the resulting Bayesian evidences are summarised in Table \ref{tab:evidences}, with the model assumptions further described in the next paragraph. In all instances, the C/O ratio was held fixed to the solar value (0.53). Wide, uniform priors were placed on all parameters except for the planetary mass, which had a Gaussian prior, with all retrieved parameters and associated priors detailed in Table \ref{tab:platon_priors}. We placed a wide prior on the temperature of the active regions to account for the effects of both spots, which are cooler than the surrounding stellar surface, and faculae/plages, which are hotter than the surrounding stellar surface. We note that when retrievals were run with stellar activity, we find that the active temperature is often lower than that of the star, indicating that the data favours unocculted spots over faculae (Tables \ref{tab:3night_table}, \ref{tab:n1_table}, \ref{tab:n2_table} and \ref{tab:n3_table}). We used nested sampling through the Python package \verb|dynesty| \citep{Speagle2020} to explore the parameter space, with 1000 live points. Nested sampling additionally gives the Bayesian evidence, allowing for robust model comparison.

\begin{table*}
\centering
\caption{The parameters and corresponding prior ranges used for PLATON retrievals. Note that not all retrievals used all parameters, see Table \ref{tab:3night_table} for further details. The values for $M_\mathrm{P}$, $R_\mathrm{P}$ ($1.263 R_\mathrm{J}$), $T_\mathrm{eq}$ ($1340K$) and $T_\mathrm{eff}$ ($5800K$) were taken from \citet{Southworth2012}. }
\label{tab:platon_priors}
\begin{tabular}{ccc}
\hline
Parameter & Units & Prior \\ \hline
Planet Mass ($M_\mathrm{P}$) & ($M_\mathrm{J}$) & Gaussian ($M_\mathrm{u} = 0.295, \sigma=0.013$) \\
Planet Radius ($R_\mathrm{P}$) & ($R_\mathrm{J}$) & Uniform ($0.8\times R_\mathrm{P}$, $1.2\times R_\mathrm{P}$) \\
Limb Temperature ($T_\mathrm{limb}$) & K & Uniform ($0.5\times T_\mathrm{eq}$, $1.5\times T_\mathrm{eq}$) \\
Metallicity ($\log Z/Z_{\odot}$) & - & Uniform (-1, 3) \\
Cloud-top Pressure ($\log P_\mathrm{cloud}$) & Pa & Uniform (-3, 7) \\
Scattering Factor ($\log s$) & - & Uniform (-2, 3) \\
Scattering Gradient ($\alpha$) & - & Uniform (-4, 30) \\
Unocculted Spot/Facula Temperature ($T_\mathrm{active}$) & K & Uniform ($T_\mathrm{eff} - 2300, T_\mathrm{eff} + 400$) \\
Unocculted Spot/Facula Covering Fraction ($f_\mathrm{active}$) & - & Uniform (0, 1) \\ 
C/O Ratio & - & Fixed at 0.53 \\ \hline
\end{tabular}
\end{table*}

To test the effects of aerosols on the scattering slope and the impacts of stellar activity, retrievals were run both with and without the two stellar activity parameters, and with the gradient of the scattering slope being both a free parameter and fixed to $\alpha$=4, equivalent to Rayleigh scattering. Of the two retrievals which account for stellar activity, the first (labelled Model 1 in Tables \ref{tab:evidences}, \ref{tab:3night_table}, \ref{tab:n1_table}, \ref{tab:n2_table}, and \ref{tab:n3_table}) included contributions from both the stellar activity and the gradient of the scattering slope ($\alpha$) as free parameters, while the second (Model 2) fitted for the contribution of active regions but held the scattering slope fixed. We refer to these models as the `free slope' and `fixed slope' models respectively, and are used to generate stellar activity correction factors detailed in Section \ref{section:activity}. Similarly, two retrievals were also performed without stellar activity, where one used a free scattering slope (Model 3) and the other kept the gradient fixed (Model 4).

\begin{figure}
    \centering
    \includegraphics[width=\linewidth]{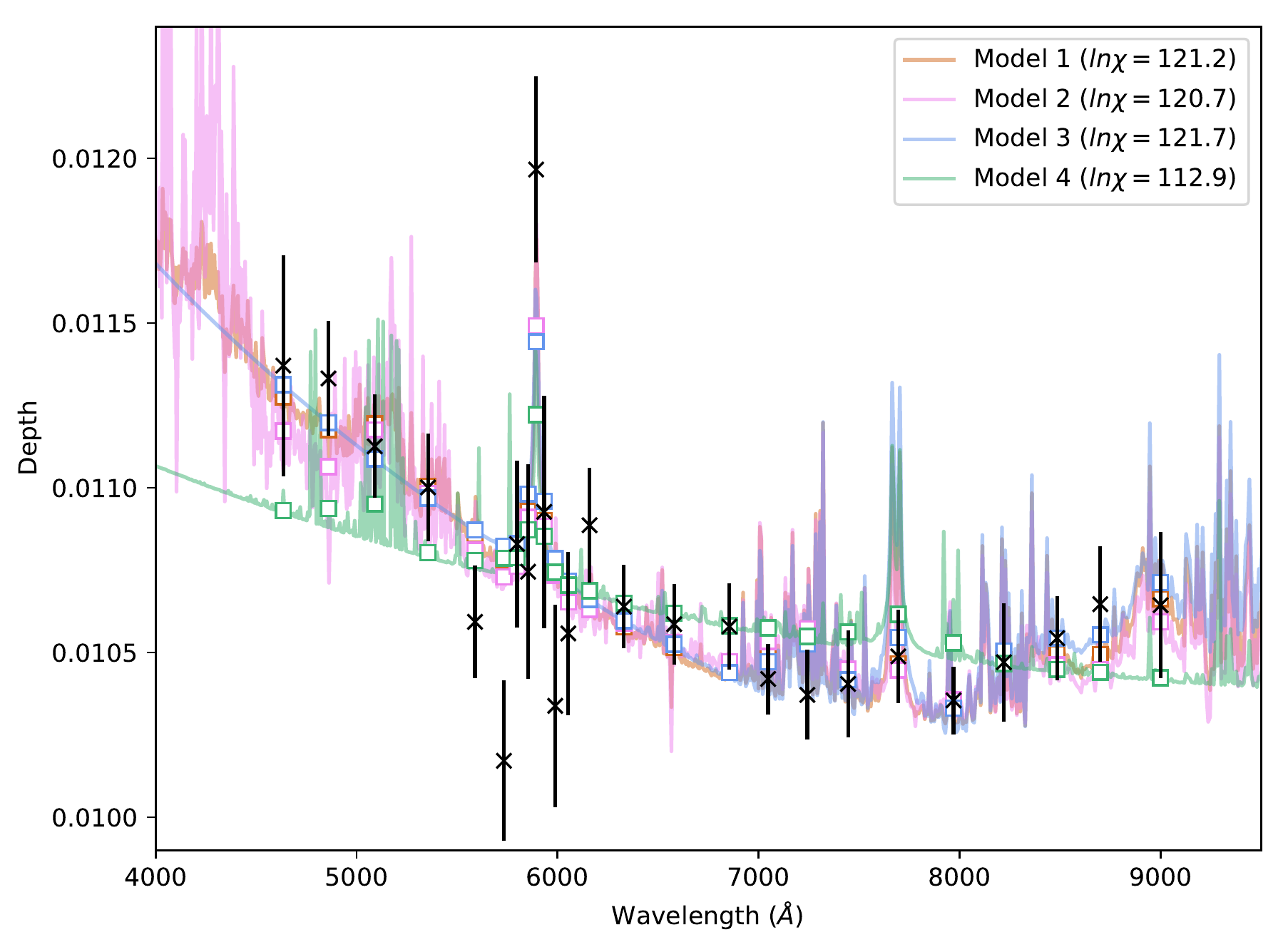}
    \caption{Atmospheric models obtained through PLATON retrievals. The black crosses correspond to our combined transmission spectrum as outlined in Table \ref{tab:3ncombined}. In each case, the line refers to an atmospheric model in Table \ref{tab:3night_table} and the matching coloured box shows the corresponding atmospheric model binned to the resolution of our black data points. The plotting in orange corresponds to Model 1, a retrieval with a free scattering slope and stellar activity. The plotting in pink corresponds to Model 2, a retrieval with $\alpha=4$ and stellar activity. The plotting in blue corresponds to Model 3, a retrieval with a free scattering slope but no stellar activity, and which is the preferred model (see Section \ref{section:activity}). The plotting in green corresponds to Model 4, a retrieval with  $\alpha=4$ and no stellar activity.}
    \label{fig:models}
\end{figure}

The results of each retrieval for the combined spectrum are listed in Table \ref{tab:3night_table}, with results from each individual night listed in the Appendix in Table \ref{tab:n1_table} for Night 1, Table \ref{tab:n2_table} for Night 2, and Table \ref{tab:n3_table} for Night 3. The retrieved model atmospheres for the combined spectrum are shown in Figure \ref{fig:models}. The nested sampling algorithm calculates the Bayesian evidences for each retrieval, given in Tables \ref{tab:3night_table}, \ref{tab:n1_table}, \ref{tab:n2_table}, and \ref{tab:n3_table} by $\ln\chi$ so as to avoid confusion with the metallicity, $\log Z$. A summary of the Bayesian evidences for all models across all data sets is given in Table \ref{tab:evidences}.


\begin{table*}
\centering
    \caption{Summary of the atmospheric models generated using PLATON and the corresponding Bayesian evidences for each data set. For further details on the retrieval results see Table \ref{tab:3night_table} for the combined transmission spectrum, and Tables \ref{tab:n1_table}, \ref{tab:n2_table} and \ref{tab:n3_table} in the Appendix for those corresponding to the individual nights.}
    \label{tab:evidences}
\begin{tabular}{cccccc}
\hline
 & \multicolumn{1}{c}{} & \multicolumn{4}{c}{Bayesian Evidences} \\
Model Name & Description & Night 1 & Night 2 & Night 3 & Combined \\ \hline
Model 1 & Free Scattering Slope \& Stellar Activity & 115.0 & 113.1 & 117.1 & 121.2 \\
Model 2 & Fixed Scattering Slope ($\alpha=4$) \& Stellar Activity & 112.6 & 113.5 & 116.5 & 120.7 \\
Model 3 & Free Scattering Slope \& No Stellar Activity & 115.0 & 111.7 & 115.9 & 121.3 \\
Model 4 & Fixed Scattering Slope ($\alpha=4$) \& No Stellar Activity & 108.1 & 110.5 & 107.3 & 112.9 \\ \hline
\end{tabular}
\end{table*}

\begin{table*}
\centering
    \caption{PLATON retrieval results for the parameters as defined in Table \ref{tab:platon_priors} for the combined transmission spectrum. Corner plots of the best fitting model (Model 3) are show in Figure \ref{fig:corner_plot} in the Appendix.}
    \label{tab:3night_table}
\begin{tabular}{ccccccccc}
\cline{2-9}
 & \multicolumn{2}{c|}{Model 1} & \multicolumn{2}{c|}{Model 2} & \multicolumn{2}{c|}{Model 3} & \multicolumn{2}{|c}{Model 4} \\ \cline{2-9} 
 & Median & Best Fit & Median & Best Fit & Median & Best Fit & Median & Best Fit \\ \hline
\multicolumn{1}{|l|}{$\ln \chi$} &  & 121.2 &  & 120.7 &  & 121.3 &  & 112.9 \\ \hline
\multicolumn{1}{|l|}{$R_\mathrm{S}$ ($R_{\odot} $)} & 1.186 (fixed) & - & - & - & - & - & - & - \\
\multicolumn{1}{|l|}{$T_\mathrm{eff}$ (K)} & 5800 (fixed) & - & - & - & - & - & - & - \\
\multicolumn{1}{|l|}{$M_\mathrm{P}$ ($M_\mathrm{J}$)} & $0.29^{+0.01}_{-0.01}$ & 0.29 & $0.29^{+0.01}_{-0.01}$ & 0.28 & $0.29^{+0.01}_{-0.01}$ & 0.29 & $0.29^{+0.01}_{-0.01}$ & 0.29 \\
\multicolumn{1}{|l|}{$R_\mathrm{P}$ ($R_\mathrm{J}$)} & $1.11^{+0.03}_{-0.04}$ & 1.11 & $1.09^{+0.02}_{-0.02}$ & 1.10 & $1.15^{+0.01}_{-0.02}$ & 1.16 & $1.13^{+0.01}_{-0.02}$ & 1.13 \\
\multicolumn{1}{|l|}{$T_\mathrm{limb}$ (K)} & $949^{+276}_{-98}$ & 825 & $867^{+97}_{-42}$ & 820 & $933^{+238}_{-68}$ & 899 & $1214^{+72}_{-136}$ & 1275 \\
\multicolumn{1}{|l|}{$\log Z$} & $0.69^{+1.23}_{-0.88}$ & -0.17  & $0.28^{+0.69}_{-0.61}$ & -0.07 & $0.39^{+0.88}_{-0.70}$ & 0.04 & $-0.04^{+0.78}_{-0.64}$ & -0.87 \\
\multicolumn{1}{|l|}{$\log P_\mathrm{cloud}$ (Pa)} & $3.68^{+1.88}_{-1.71}$ & 4.86 & $4.65^{+1.40}_{-1.25}$ & 4.71 & $4.26^{+1.67}_{-1.41}$ & 5.84 & $4.64^{+1.46}_{-1.23}$ & 3.54 \\
\multicolumn{1}{|l|}{$\log s$} & $-0.81^{+1.43}_{-0.79}$ & -1.68 & $-0.21^{+0.76}_{-0.84}$ & -0.33 & $-1.25^{+0.72}_{-0.47}$ & -1.67 & $1.50^{+0.49}_{-0.44}$ & 1.11 \\
\multicolumn{1}{|l|}{$\alpha$} & $11.8^{+4.6}_{-5.9}$ & 11.1 & 4 (fixed) & - & $13.9^{+2.5}_{-2.3}$ & 13.8 & 4 (fixed) & - \\
\multicolumn{1}{|l|}{$T_\mathrm{spot}$ (K)} & $5239^{+460}_{-780}$ & 4632 & $5154^{+362}_{-523}$ & 5578 & $T_\mathrm{eff}$ (fixed) & - & - & - \\
\multicolumn{1}{|l|}{$f_\mathrm{active}$} & $0.22^{+0.26}_{-0.12}$ & 0.15 & $0.33^{+0.32}_{-0.10}$ & 0.81 & 0 (fixed) & - & - & - \\
\multicolumn{1}{|l|}{C/O} & 0.53 (fixed) & - & - & - & - & - & - & - \\ \hline
\end{tabular}
\end{table*}

In all cases, our retrieval results show that models with a fixed scattering slope without the inclusion of stellar activity (Model 4 in Tables \ref{tab:3night_table}, \ref{tab:n1_table}, \ref{tab:n2_table} and \ref{tab:n3_table}) are least favoured, and we therefore rule out this combination for WASP-21b. For the rest of the models, the values of $\ln\chi$ are similar within each individual night's spectra, however there is some variation in the most favoured model between the nights (Tables  \ref{tab:n1_table}, \ref{tab:n2_table} and \ref{tab:n3_table}). Despite this, the retrieved parameters are similar across each of the individual nights, and with the combined transmission spectrum. Higher values of $\ln\chi$ are obtained for the models retrieved using the combined spectrum (Table \ref{tab:3night_table}), and we therefore use this spectrum to rule out further models. The consistency of the retrievals both between nights and with the combined spectrum demonstrates the validity of combining the three nights to form a single spectrum, a decision which is further supported by the good agreement between the transmission spectra of each of the individual nights (Figure \ref{fig:trans_spec}, First Panel). 

Given that all retrieved values of the limb temperature are significantly cooler than the equilibrium temperature of WASP-21b, retrievals were also run with the temperature held fixed to the equilibrium temperature. However, such retrievals led to poor fits, and so we do not include them here. We note that because \verb|PLATON| retrieves the temperature of the limb of the atmosphere, such discrepancies are likely. Cold retrieved temperatures are not uncommon, and 1D atmospheric retrieval techniques have been shown to be biased towards cooler temperatures \citep{Macdonald2020}.



For the combined transmission spectrum, `free slope' models (Models 1 and 3 in Table \ref{tab:3night_table}) are marginally favoured over the remaining `fixed slope' model (Model 2). In both free slope models, large scattering slope gradients are retrieved, with median values of $\alpha=11.8^{+4.6}_{-5.9}$ when including stellar activity, and $\alpha=13.9^{+2.5}_{-2.3}$ without. Between the `free slope' models, while the retrieval without stellar activity has the higher value of $\ln\chi$, the Bayesian evidences alone cannot clearly identify a favoured model, with $\Delta\ln\chi=0.1$. Given the conclusions of Section \ref{section:photometry} that WASP-21 is an inactive star, Model 3, a retrieval which considers a free scattering slope but no stellar activity, is our favoured interpretation of the atmosphere of WASP-21b over Models 1 and 2. However, given the potential for stellar activity to mimic atmospheric features, the possible impacts of stellar activity are discussed in more detail in Section \ref{section:activity}.

\section{Discussion} \label{section:discussion}

\begin{figure*}
    \centering
    \includegraphics[width=\linewidth]{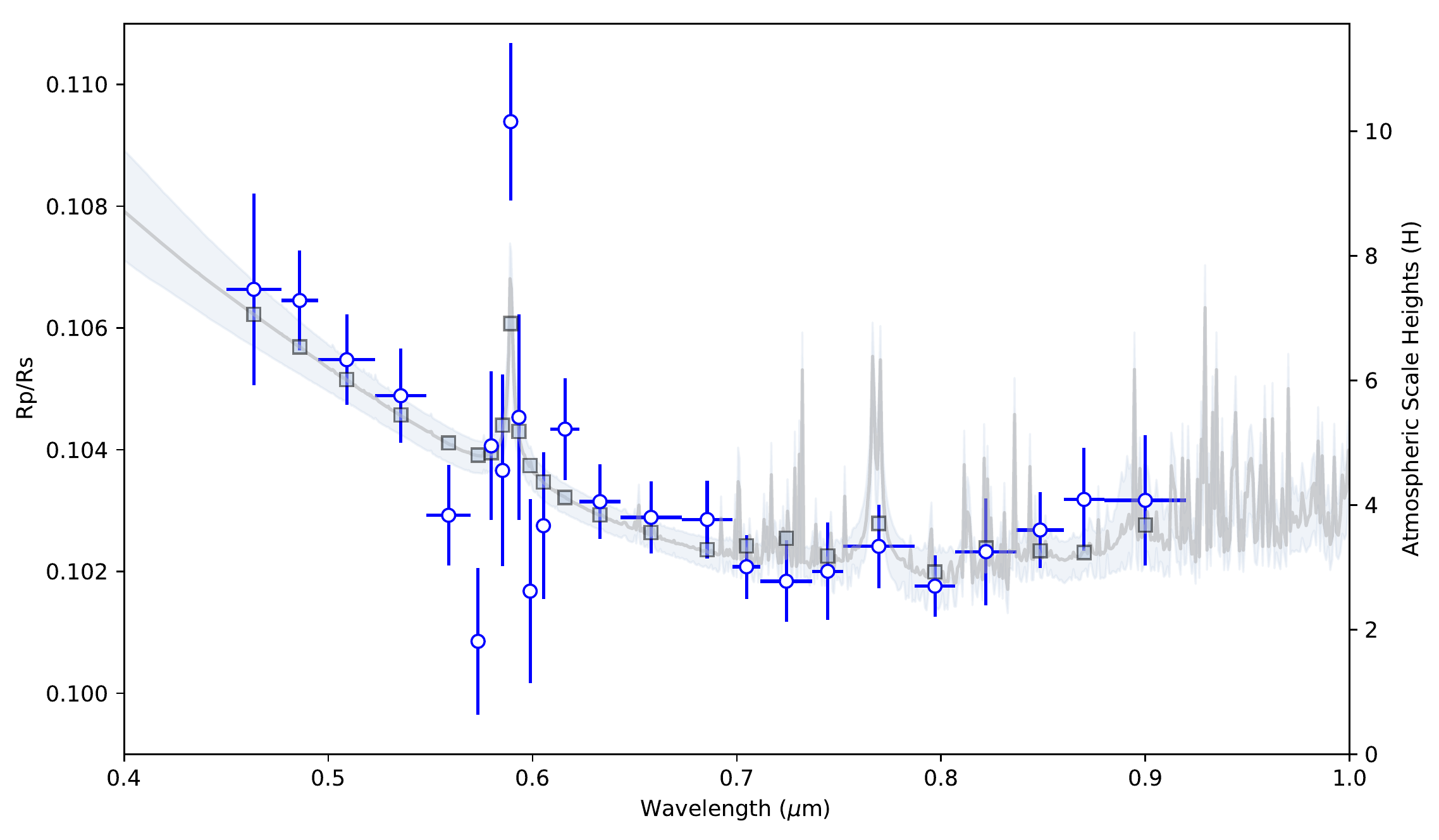}
    \caption{Our best fitting atmospheric model retrieved using PLATON \citep{Zhang2019}. The blue points correspond to our final combined transmission spectrum (Table \ref{tab:3ncombined}). The grey line shows the unbinned best fitting atmospheric model while the grey boxes give the model binned to the resolution of the blue data points. The shaded region corresponds to the $1\sigma$ uncertainty in the model.}
    \label{fig:final_spec}
\end{figure*}

\subsection{On the Impact of Stellar Activity} \label{section:activity}

To determine whether the features observed in the transmission spectrum of WASP-21b are present in the atmosphere, or induced by potential stellar activity, we corrected both our combined and individual nights' spectra using wavelength-dependent activity correction factors generated during retrievals by \verb|PLATON|. Correction factors were obtained for all retrievals which take into account stellar activity (Models 1 and 2), leading to both a `free slope' and a `fixed slope' correction for each data set. Dividing the transmission spectrum by these correction factors allows the contribution of the planet to be separated and isolated, highlighting whether features are truly present in the atmosphere.

Confidence levels of the detection of the \ion{Na}{i} feature were recalculated for the combined and individual nights' spectra using correction factors for both the `free' and `fixed' models, as listed in Table \ref{tab:na_sig}. While in all cases the significance of the detection decreases after the application of the stellar activity corrections, the combined spectrum still indicates a greater than 3$\sigma$ detection of \ion{Na}{i} absorption irrespective of any activity correction. These results imply that stellar activity cannot be the single source of the observed absorption features, in agreement with the low levels of activity seen in WASP-21, as outlined in Section \ref{section:photometry}.

As unocculted spots and occulted plages are capable of inducing blueward slopes which can mimic Rayleigh scattering slopes \citep{McCullough2014, Oshagh2014}, it is necessary to also consider whether the observed slope in the transmission spectrum of WASP-21b could be caused by stellar activity. As shown by Figure \ref{fig:models} and by the similarity of the scattering slope gradients within $1\sigma$ uncertainties, the inclusion or exclusion of stellar activity in retrievals with free scattering slopes appears to have no significant impact on the retrieved scattering slope parameters (Table \ref{tab:3night_table}). While the retrieval with the scattering slope gradient fixed at $\alpha=4$ including stellar activity results in an almost comparable fit to the data ($\Delta\ln\chi=0.5$ and 0.6) to those with free scattering slopes, large values of $f_\mathrm{active}$ are retrieved. In addition, when $\alpha=4$, the amplitude of the stellar absorption features increases significantly when compared to Model 1, which has a free scattering slope (Fig. \ref{fig:models}).

\begin{figure}
    \centering
    \includegraphics[width=\linewidth]{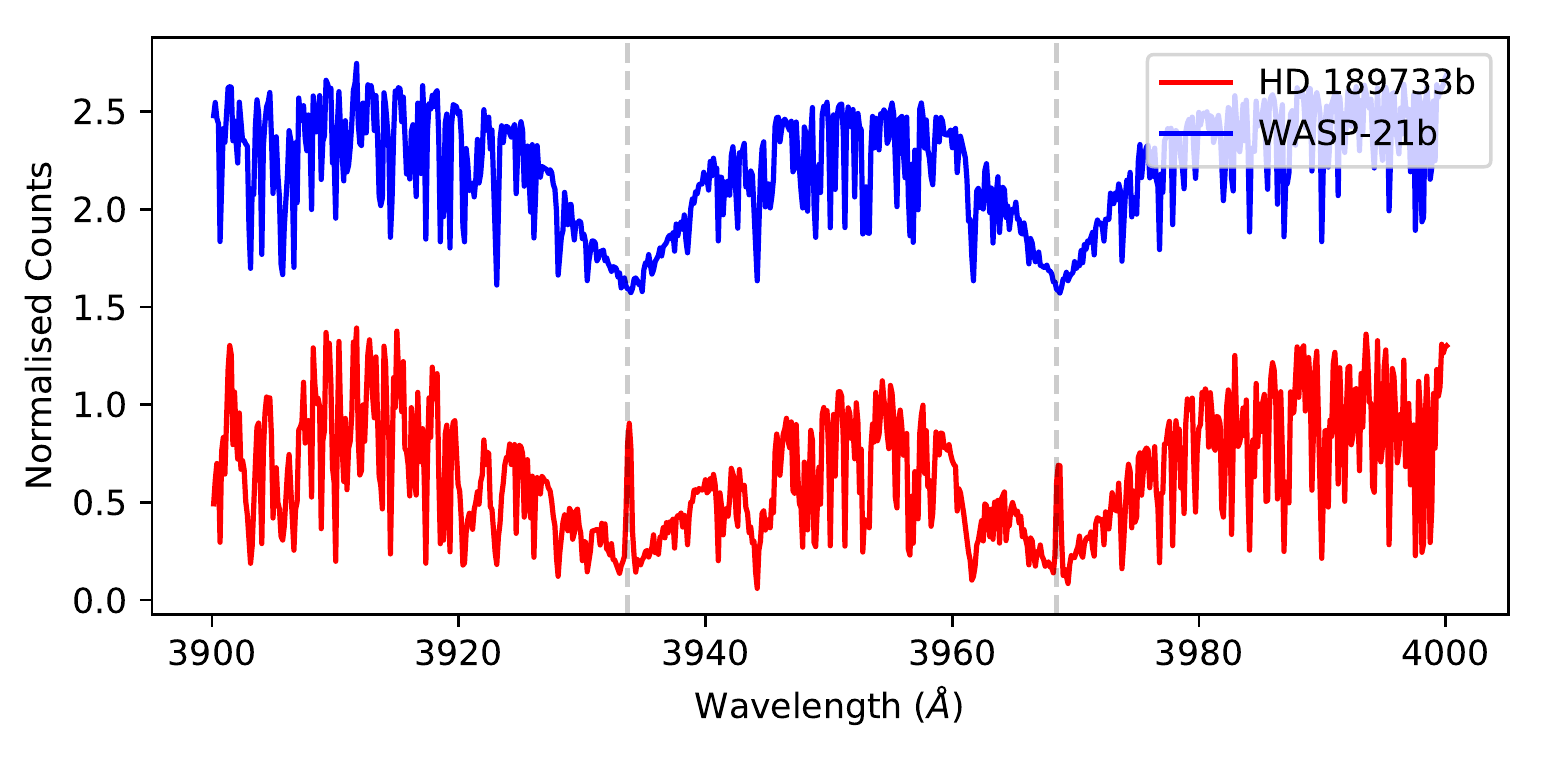}
    \caption{Spectra of WASP-21 (blue line) and HD\,189733 (red line) obtained with the HARPS spectrograph \citep{Mayor2003} with the positions of the \ion{Ca}{ii} H\&K lines at 3968.47\,\AA\, and 3933.66\,\AA\, respectively.}
    \label{fig:cahk}
\end{figure}

While the low photometric amplitude observed in Section \ref{section:photometry} implies that WASP-21 does not likely have a significant coverage of unocculted spots, stellar faculae and plages can produce lower photometric variability and are therefore less likely to be detected through photometric monitoring \citep[e.g.][]{Shapiro2016,Morris2018}. Occulted plages and faculae are also capable of inducing blueward slopes in transmission spectra \citep{Oshagh2014}, and may be difficult to detect if they cover a significant fraction of the stellar disk but have only a small temperature contrast. 

In the specific case of HD\,189733b, \citet{Oshagh2014} find that the observed transmission spectrum can be reproduced with a plage covering factor of 1.96\% given a temperature contrast of 100\,K. Since we find a scattering slope very similar to that of HD\,189733b (see Section \ref{section:hd189}), we consider whether or not occulted plages could lead to a similar effect in WASP-21b. Large covering fractions of plages and faculae would lead to significant core emission in the \ion{Ca}{ii} H\&K lines, which can be assessed using high resolution spectroscopy \citep[e.g.][]{Berdyugina2005, Knutson2010, Morris2018}. Using spectra obtained from observations made with the HARPS instrument at La Silla \citep{Mayor2003}, we were able to compare the \ion{Ca}{ii} H\&K lines of WASP-21 to that of HD\,189733, which \citet{Knutson2010} show to have very evident \ion{Ca}{ii} H\&K line core emission, reflective of an active star. Figure \ref{fig:cahk} shows that compared to HD\,189733, WASP-21 has no significant \ion{Ca}{ii} H\&K line core emission. A lack of emission is indicative of a less active star, and it is therefore unlikely that WASP-21 could have a covering fraction of faculae and plages similar to that postulated for HD\,189733. While a very small and bright occulted plage or facula could lead to a slope in the transmission spectrum and not necessarily cause emission in the line cores of \ion{Ca}{i} H\&K, such a feature would generate a discrete dip in a transit light curve, and is not seen for WASP-21b (Figures \ref{fig:wl_curves}, \ref{fig:n1_sl_curves}, \ref{fig:n2_sl_curves}, \ref{fig:n3_sl_curves}). WASP-21 is also an older star ($12\pm5$ Gyr, \citealt{Bouchy2010}) than HD\,189733 ($6.8^{+5.2}_{-4.4}$ Gyr \citealt{Torres2008}) and while these ages have large uncertainties, stellar age-activity relations \citep[e.g.][]{Berdyugina2005, Lorenzo2016} reflect that the older WASP-21 is likely less active than HD\,189733.

In addition, \citet{Rackham2019} show that G3V type stars such as WASP-21 typically have a spot coverage of $0.4^{+0.8}_{-0.2}$\%, meaning that among G3V stars WASP-21 would be a $1.7\sigma$ outlier in the case of Model 1 ($f_\mathrm{active}=22^{+26}_{-12}\%$) and a $3\sigma$ outlier in the case of Model 2 ($f_\mathrm{active}=33^{+32}_{-10}\%$) (Table \ref{tab:3night_table}). While Model 1 is not a significant outlier, using the empirical scaling relation between photometric modulation amplitude and covering fraction, and scaling relation coefficients from \citet{Rackham2019}, the 22\% covering fraction of Model 1 would give a $2.5\pm1.2\%$ amplitude change in the photometry. This is not consistent with the results obtained through the AIT data outlined in Section \ref{section:photometry}, and is further evidence that WASP-21 does not have significant levels of activity.


Collectively, these retrieval results, along with the conclusions of the photometric monitoring described in Section \ref{section:photometry} and the lack of \ion{Ca}{ii} H\&K emission, imply that is is unlikely that stellar activity plays a significant contribution to the features observed in the transmission spectrum presented in this work. In particular, we find it likely that the scattering slope observed in the transmission spectrum of WASP-21b is of planetary origin and not induced by unocculted spots or occulted faculae. Therefore, we conclude that the atmosphere of WASP-21b features both a scattering slope and \ion{Na}{i} absorption, and is best fit by an atmospheric retrieval with a free scattering slope without the inclusion of stellar activity (Model 3, Table \ref{tab:3night_table}), as presented in Figure \ref{fig:final_spec}.

\subsection{The Detection of Sodium Absorption} \label{section:na_dis}

As detailed in Section \ref{section:sodium}, our combined transmission spectrum of WASP-21b shows a detection of \ion{Na}{i} absorption extending over 5 scale heights at 4.03$\sigma$ confidence (Fig. \ref{fig:trans_spec}, Table \ref{tab:na_sig}), which extends to 100\,\AA\, (Fig. \ref{fig:iib}). Low-resolution ground-based detection of the \ion{Na}{i} feature are rare, and previously have only been made in a handful of exoplanets, XO-2b \citep{Sing2012}, WASP-39b \citep{Nikolov2016}, WASP-19b \citep{Sedaghati2017}, WASP-52b \citep{Chen2017}, WASP-127b \citep{Chen2018}, and WASP-96b \citep{Nikolov2018}. The \ion{Na}{i} feature in WASP-21b extends over an equal number of scale heights to several other exoplanets (e.g. HAT-P-1b, \citealt{Nikolov2013}; WASP-39b, \citealt{Nikolov2016}; WASP-19b, \citealt{Sedaghati2017}), however we note that the significance of the detection in WASP-21b is larger than that of these three planets. 

The two most prominent existing cases of \ion{Na}{i} absorption are that of WASP-39b \citep{Sing2016, Fischer2016a, Nikolov2016} and WASP-96b \citep{Nikolov2018}. In both of these exoplanets, the pressure broadened wings of the \ion{Na}{i} feature are detected, with the wings extending out to $\sim1000$\,\AA, and are taken as evidence that the atmospheres of these planets are relatively cloud free. Detections of the predicted broad absorption features in exoplanet atmospheres are notably uncommon, with both ground and space-based observations of the \ion{Na}{i} resonance doublet typically only revealing the narrow core of the feature \citep[e.g.][]{Charbonneau2002, Snellen2008, Redfield2008, Huitson2012, Nikolov2013, Wilson2015, Carter2019, Chen2020}. In the case of WASP-21b, the narrower feature shown in Figures \ref{fig:trans_spec} and \ref{fig:iib} indicates that aerosols could be preventing this feature from being as broad as that of WASP-96b and WASP-39b, which would be consistent with the observed scattering slope. We discuss the implication of the presence of aerosols, particularly in the context of the steep scattering slope in further detail in Section \ref{section:hd189}.

Given the theoretical predictions for broad absorption features from both \ion{Na}{i} and \ion{K}{i} \citep{Seager2000, Brown2001}, and the strength of the detection of \ion{Na}{i} in the atmosphere of WASP-21b, the apparent lack of \ion{K}{i} (Fig. \ref{fig:trans_spec}) poses an interesting point of discussion. As demonstrated by Figures \ref{fig:bins} and \ref{fig:trans_spec}, the centre of the \ion{K}{i} resonance doublet lies $\sim$60\,\AA\, redward of the telluric oxygen feature, and therefore observations of \ion{K}{i} are particularly difficult from the ground. We note that the 30\,\AA\, bin shown in the middle panel of Figure \ref{fig:trans_spec} is centred on the right hand edge of the \ion{K}{i} doublet due to proximity of this telluric feature, and so only covers the right-hand line. We therefore also tested a wider bin covering both the left-hand and right-hand lines, but did not obtain a significantly deeper signal. As the precision of the transit depth in the continuum around the \ion{K}{i} feature is comparable to the rest of the transmission spectrum, we have confidence that the telluric O2 A band feature is not having a significant impact on our transmission spectrum. It is therefore likely that the atmosphere of WASP-21b is depleted in potassium, and adds to the growing number of exoplanets that have only one alkali metal present in their atmospheres. While the exact cause of this alkali discrepancy is currently unknown, postulated theories include primordial abundance variation, the potential photoionisation of potassium by stellar UV radiation from the host, and the formation of condensates such as potassium chloride within the atmosphere, all of which would lead to depleted line cores \citep{Nikolov2018}. Photoionisation favours the depletion of K over Na due to the lower photoionisation of K \citep{fortney2003, Sing2015}. However the difference in the photoionisation energies is only 0.8\,eV, and so we might expect Na to also be depleted if photoionisation is the cause \citep{Barman2007}. If the alkali metals are condensing into the chlorides NaCl and KCl, this could reduce the upper atmosphere of atomic Na and K \citep[e.g.][]{Sing2008,Wakeford2015}. However, the condensation temperature of KCl is only 85\,K lower than that of NaCl at a pressure of 1\,mbar \citep{Palik1998, Wakeford2015} and so we might expect this to preferentially remove Na and not K. As a result, we believe that a primordial abundance variation is the most likely cause of the observed imbalance.

While this manuscript was under review, an independent study by \citet{Chen2020} published a detection of \ion{Na}{i} in the atmosphere of WASP-21b, confirming our finding.

\subsection{Contextualising the Scattering Slope of WASP-21b} \label{section:hd189}

While the presence of a deep \ion{Na}{i} absorption feature indicates that WASP-21b may have a cloud-free atmosphere, its strong scattering slope ($\alpha=13.9$, Table \ref{tab:3night_table}) implies that aerosols are likely present \citep{Wakeford2015, Pinhas2017}. In order to put this steep slope into context, we compare the transmission spectrum of WASP-21b to that of HD\,189733b \citep{Bouchy2005}, using the transmission spectrum of \cite{Pont2013} obtained with the Hubble Space Telescope (HST). HD\,189733b is a hot Jupiter known to possess an atmosphere with a significant scattering slope caused by the presence of aerosols, such that only the narrow \ion{Na}{i} absorption feature is observed \citep{Pont2008, LecavelierDesEtangs2008, Sing2011a, Huitson2012, Pont2013}. For these reasons, HD\,189733b provides a useful comparison to the features observed in WASP-21b.

\begin{figure}
    \centering
    \includegraphics[width=\linewidth]{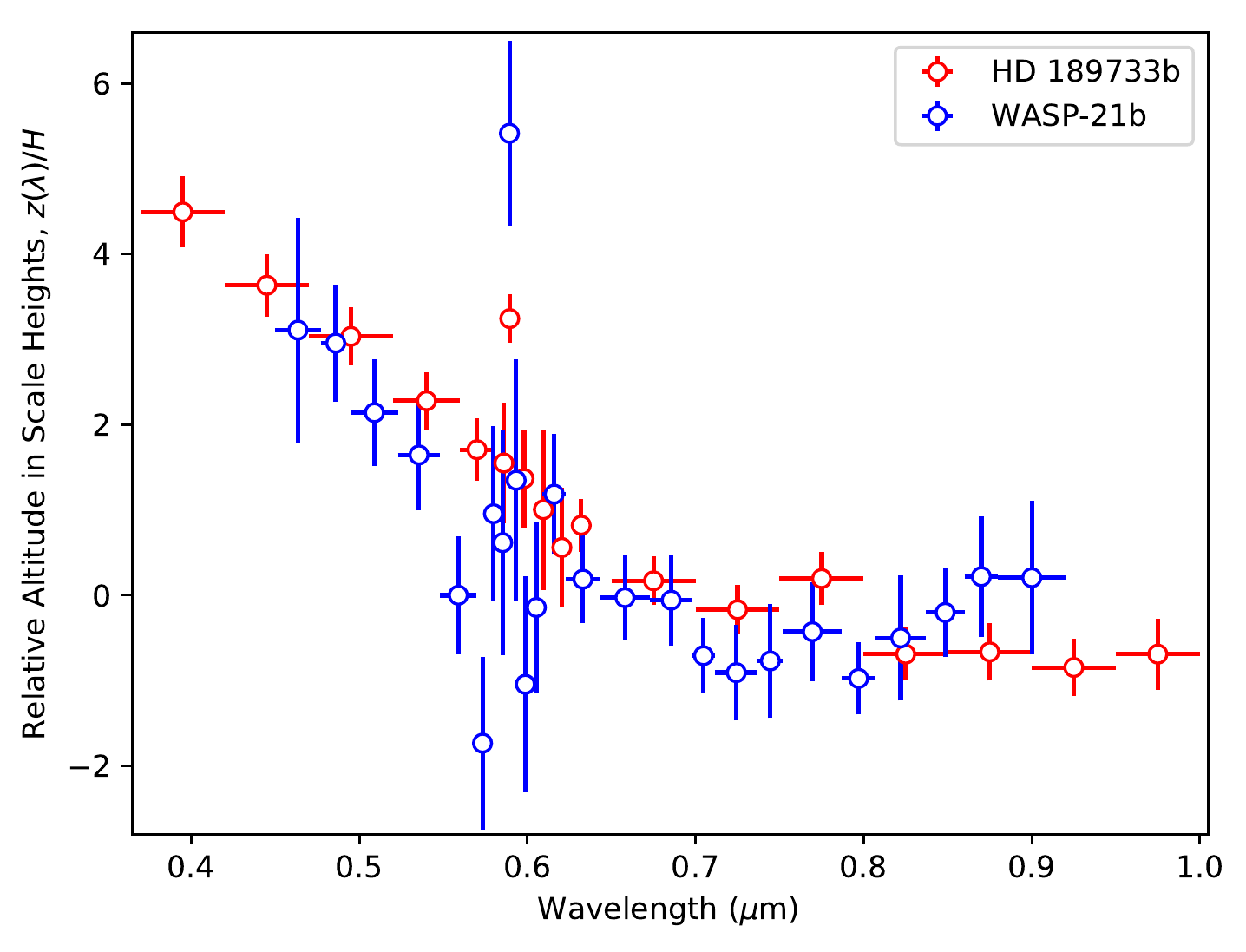}
    \caption{Our transmission spectrum of WASP-21b (blue circles) and the HST transmission spectrum of HD\,189733b from \citet{Pont2013} (red circles), plotted as a function of their respective atmospheric scale heights.}
    \label{fig:hd198}
\end{figure}

Figure \ref{fig:hd198} shows the transmission spectra of WASP-21b and HD\,189733b plotted as a function of their respective atmospheric scale heights. WASP-21b is seen to have a very similar blueward slope to HD\,189733b, however, notably WASP-21b has a deeper \ion{Na}{i} absorption feature than HD\,189733b, extending over 5 scale heights in WASP-21b, compared to only 2 in HD\,189733b. WASP-21b also shows signs of a redward rise in its transmission spectrum where HD\,189733b remains primarily featureless \citep{Pont2013}.  

WASP-21b has a slightly higher equilibrium temperature than HD\,189733b, with WASP-21b having T$_\mathrm{eq}=1340$\,K \citep{Southworth2012} and HD\,189733b having T$_\mathrm{eq}=1191$\,K \citep{Southworth2010}. Equilibrium temperature has been predicted to influence the degree of cloudiness in a planet's atmosphere \citep[e.g.][]{Stevenson2016, Heng2016, Fu2017,Crossfield2017,Evans2018,Hoeijmakers2018,Hoeijmakers2019}, with hotter exoplanets more likely to be cloud-free. This would suggest that WASP-21b could have a clearer atmosphere than HD\,189733b, however there is some evidence to suggest that higher surface gravities may also lead to clearer atmospheres \citep{Stevenson2016}, and WASP-21b ($g=5.07$\,ms$^{-2}$, \citealt{Ciceri2013}) has a significantly lower surface gravity than HD\,189733b ($g=21.5$\,ms$^{-2}$, \citealt{Southworth2010}). It therefore may be possible that these two parameters have contrasting effects in the case of HD\,189733b and WASP-21b, causing the two planets to have similar aerosol levels despite their different properties, as demonstrated by the good agreement in their scattering slopes in Figure \ref{fig:hd198}.

Despite the similarities between WASP-21b and HD\,189733b, it is difficult to definitively say what may be causing either of their scattering slopes. There is currently no clear consensus as to what causes the enhanced scattering slope of HD\,189733b. Postulated theories range from a high altitude haze \citep{Sing2016, Barstow2020}, mineral clouds consisting of small particles such as MgSiO$_3$ and SiO$_2$ \citep{LecavelierDesEtangs2008, Pont2013, Helling2016} or a combination of cloud and haze layers \citep{Pont2013, Pinhas2019}, however it is clear that no single species can explain the slope \citep{Wakeford2015, Pinhas2017, Helling2016, Barstow2020}. Unocculted spots and occulted plages have also been proposed as the potential cause of the slope in HD\,189733b \citep{McCullough2014, Oshagh2014}, as we discuss in the case of WASP-21b in Section \ref{section:activity}. However, in the case of unocculted spots for HD\,189733b, the required spot coverage is more substantial than that estimated by \cite{Pont2013}. The fact that WASP-21b has been shown to have an inactive host (see Sections \ref{section:photometry} and \ref{section:activity}) demonstrates that scattering slopes as steep as that observed in the atmospheres of WASP-21b and HD\,189733b can occur without the influence of stellar activity. The scattering slope of WASP-21b could be caused by any one, or a combination of the explanations outlined for HD\,189733b, although as discussed in Section \ref{section:activity}, we do not believe stellar activity is the cause. Indeed, a recent study of aerosol composition in giant exoplanets by \cite{Gao2020} found that silicate species dominate aerosol opacities in exoplanets with $T_\mathrm{eq}>950K$ due to their low nucleation energy barriers and high elemental abundances. Such findings are further evidence that the presence of aerosols, particularly silicates like Mg$_{2}$SiO$_{4}$ and MgSiO$_{3}$, are the cause of the scattering slope in WASP-21b. Future infra-red observations of WASP-21b would be able to further constrain the degree to which aerosols are present within its atmosphere, and potentially identify the molecules present, clarifying the causes of these steep scattering slopes.

Another interesting comparison to WASP-21b is WASP-39b \citep{Faedi2011}, which has a near identical mass to WASP-21b at 0.28\,M$_\mathrm{Jup}$ \citep{Mancini2018} compared to WASP-21b's 0.30\,M$_\mathrm{Jup}$ \citep{Southworth2012}. While WASP-21b has a higher equilibrium temperature than WASP-39b (T$_\mathrm{eq}=1166$\,K \citealt{Mancini2018}), WASP-39b has one of the clearest atmospheres of any known exoplanet \citep{Sing2016, Fischer2016a, Nikolov2016, Wakeford2018, Kirk2019}, suggesting that temperature alone cannot predict the cloudiness of an atmosphere. WASP-39b has had its metallicity measured by several groups who have reported significantly different metallicities \citep{Tsiaras2018, Fisher2018, Wakeford2018, Kirk2019, Pinhas2019, Thorngren2019, Welbanks2019}, which has implications for the mass-metallicity relation of giant exoplanets. As the mass of WASP-21b is so similar to that of WASP-39b, measurements of its water abundance to determine its atmospheric metallicity would provide an excellent test of the mass-metallicity relation in the Saturn-mass regime. Future HST observations of the water absorption feature at 1.4 microns would not only allow the mass-metallicity relation to be tested, but would also provide further constraints on the cloudiness of the atmosphere of WASP-21b. 

\section{Conclusions} \label{section:conclusion}

We present the study of the atmosphere of the hot Saturn WASP-21b. Our ground-based optical transmission spectrum has been obtained through the LRG-BEASTS survey, which continues to demonstrate that 4-metre class telescopes can obtain transmission spectra with precisions comparable to that of 8 and 10-metre class telescopes. The combined transmission spectrum obtained covers a wavelength range of 4635-9000\,\AA\, at an average precision of 197\,ppm.

WASP-21b has been shown to have a strong scattering slope best fit by a gradient of $13.9^{+2.5}_{-2.3}$ (compared to 4 for Rayleigh scattering), along with a detection of sodium absorption at 4.03$\sigma$ confidence in a 30\,\AA\, wide bin. We see no evidence of potassium absorption in a 30\,\AA\, wide bin. Detections of \ion{Na}{i} of this significance are rare from low-resolution ground-based observations, and add WASP-21b to the growing list of exoplanets with evidence of only one alkali metal present in their atmospheres.

Stellar activity is not found to be a significant contributor to the features observed in the transmission spectrum, and cannot alone explain the detection of the \ion{Na}{i} feature or the scattering slope. The lack of photometric modulation and absence of \ion{Ca}{ii} H\&K line core emission indicate that unocculted spots and occulted faculae/plages are not responsible for the steep slope we observed in WASP-21b's transmission spectrum. 


While a strong \ion{Na}{i} detection at low-resolution can indicate a clear atmosphere, the gradient of the scattering slope observed is similar to that of HD\,189733b, and so we conclude that WASP-21b likely has aerosols present in its atmosphere. Future observations of the water absorption feature at 1.4 microns would clarify whether WASP-21b is truly similar to HD\,189733b.

\section*{Acknowledgements}

We thank the anonymous referee for their helpful suggestions and comments. We also thank Munazza Alam, Victoria DiTomasso and Ian Weaver for useful and insightful discussions, and Benjamin Rackham for helpful discussions regarding stellar activity. LA thanks the support of the University of Southampton (Astrophysics with a Year Abroad programme) and the Center for Astrophysics | Harvard \& Smithsonian, where this research was undertaken. PJW, TL and GK have been supported by STFC consolidated grant ST/P000495/1. PJW and MB are also supported by ST/T000406/1. GWH acknowledges long-term support from NASA, NSF, Tennessee State University, and the State of Tennessee through its Centers of Excellence program. The WHT is operated on the island of La Palma by the Isaac Newton Group of Telescopes in the Spanish Observatorio del Roque de los Muchachos of the Instituto de Astrofísica de Canarias. Based on data obtained from the ESO Science Archive Facility under request number 559101. 

\section*{Data Availability}

The reduced light curves presented in this work will be made available with the online journal version of this paper and at the CDS (http://cdsarc.u-strasbg.fr/)



\bibliographystyle{mnras}
\bibliography{refs}



\appendix

\section{Additional Figures}

\begin{figure*}
    \centering
    \includegraphics[height=20cm]{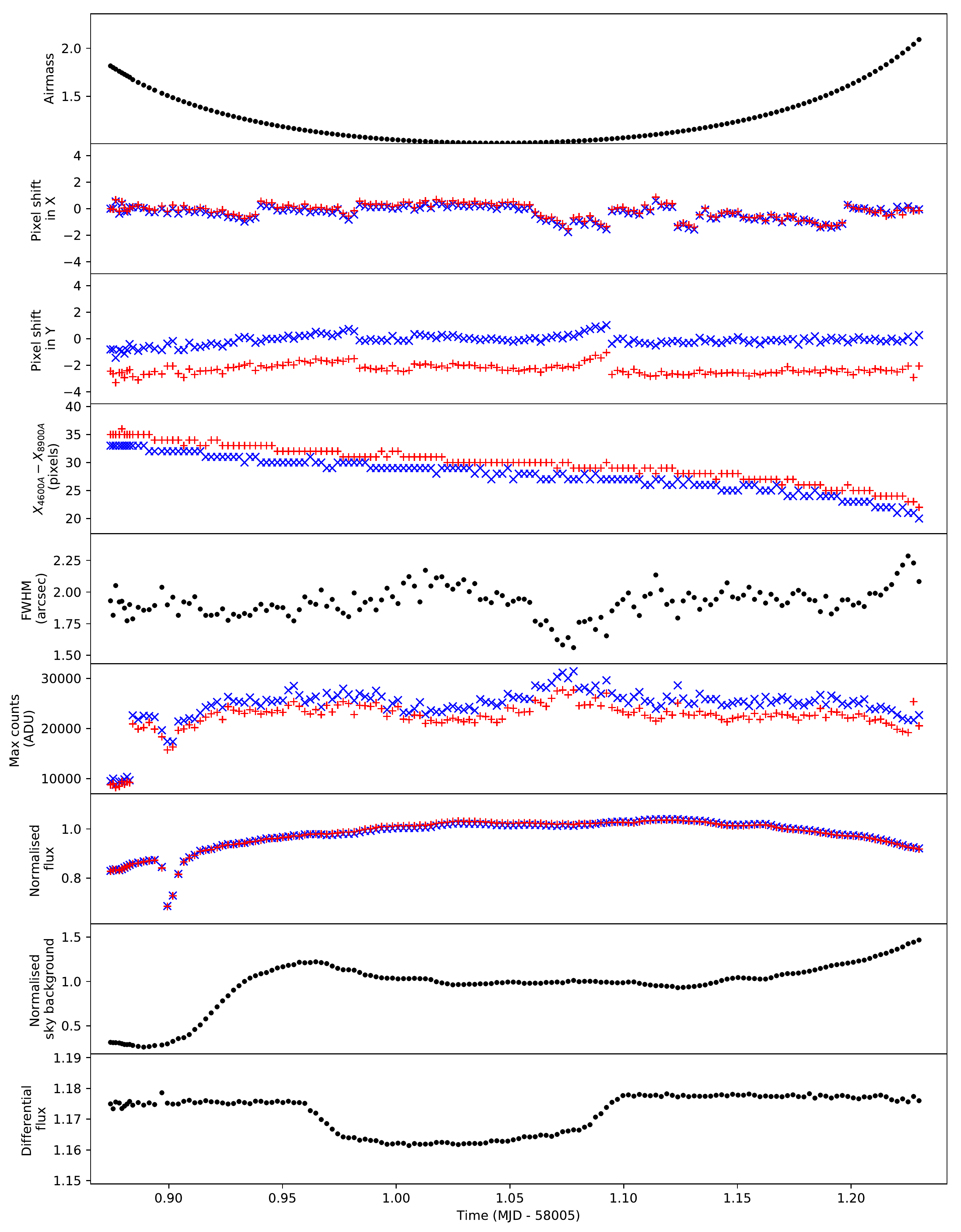}
    \caption{Diagnostic plots of the first night's data. In all panels time is plotted on the $x$-axis. Top panel: variation of airmass throughout the observations. Second panel: the shift in the spatial direction in the target's (blue crosses) and comparison's (red pluses) spectra. Third panel: the shift in the dispersion direction in the target's (blue crosses) and comparison's (red pluses) spectra. Fourth panel: rotation of the target's (blue crosses) and comparison's (red pluses) spectra as measured by the difference in the traces' locations at red and blue wavelengths. Fifth panel: variation in the FWHM throughout the observations. Sixth panel: maximum counts recorded in the target's (blue crosses) and comparison's (red pluses) spectra. Seventh panel: the raw white-light curves of the target (blue crosses) and comparison (red pluses). Eighth panel: normalised sky background throughout the observations. Bottom panel: WASP-21's white light curve following division by the comparison's light curve.}
    \label{fig:n1_ancillary}
\end{figure*}

\begin{figure*}
    \centering
    \includegraphics[height=20cm]{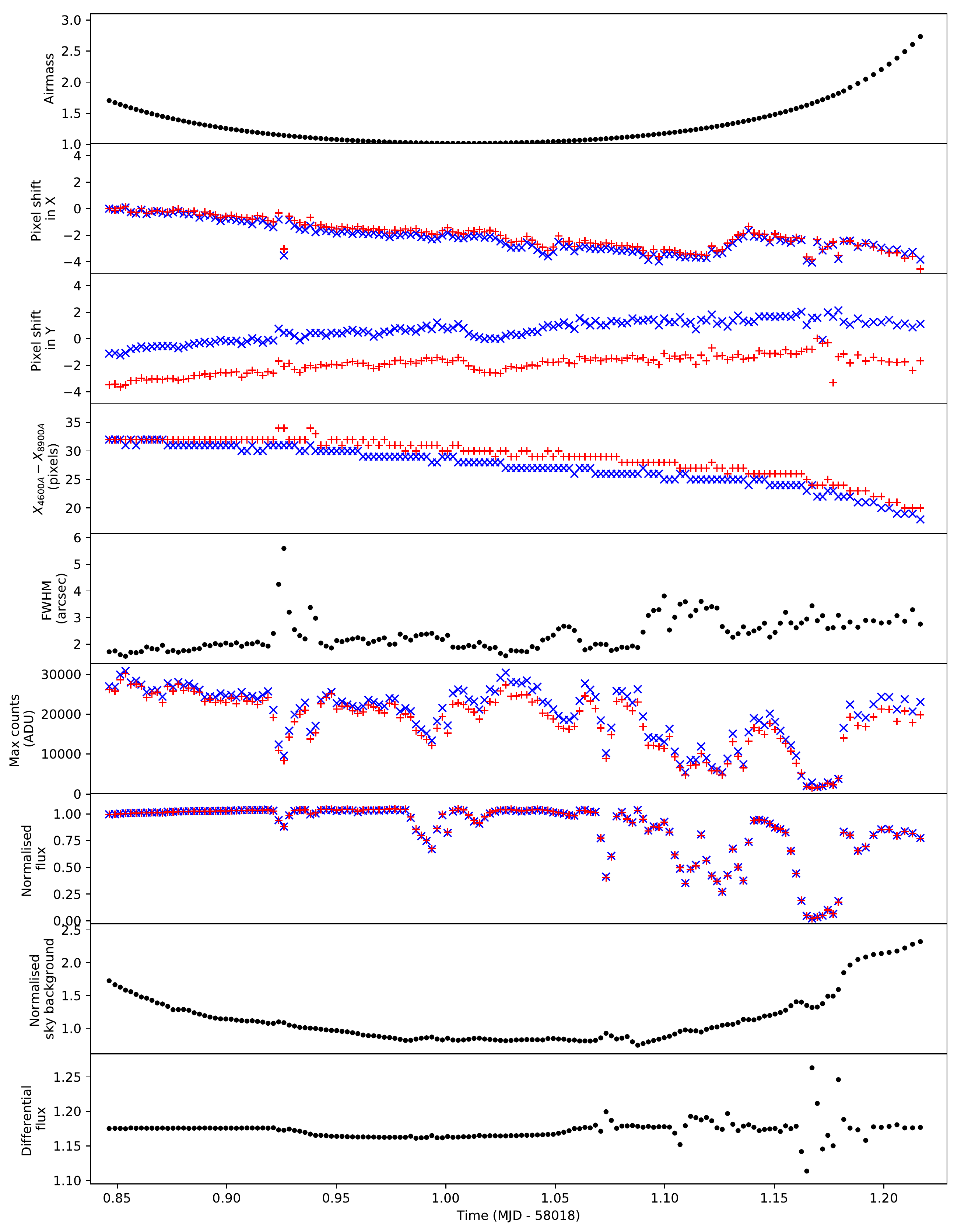}
    \caption{Diagnostic plots of the second night's data. See Figure \ref{fig:n1_ancillary} for details.}
    \label{fig:n2_ancillary}
\end{figure*}

\begin{figure*}
    \centering
    \includegraphics[height=20cm]{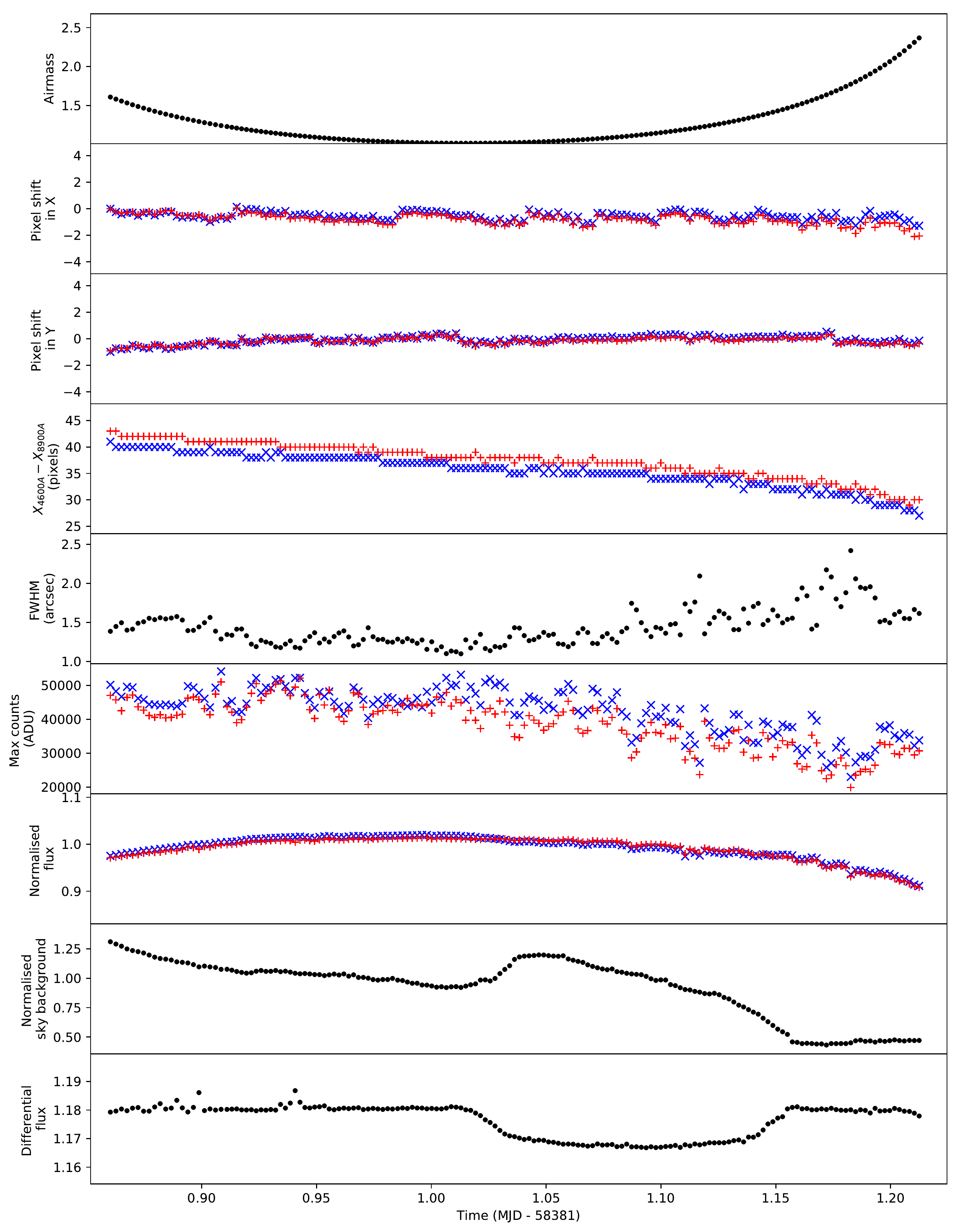}
    \caption{Diagnostic plots of the third night's data. See Figure \ref{fig:n1_ancillary} for details.}
    \label{fig:n3_ancillary}
\end{figure*}

\begin{figure*}
    \centering
    \includegraphics[width=\linewidth]{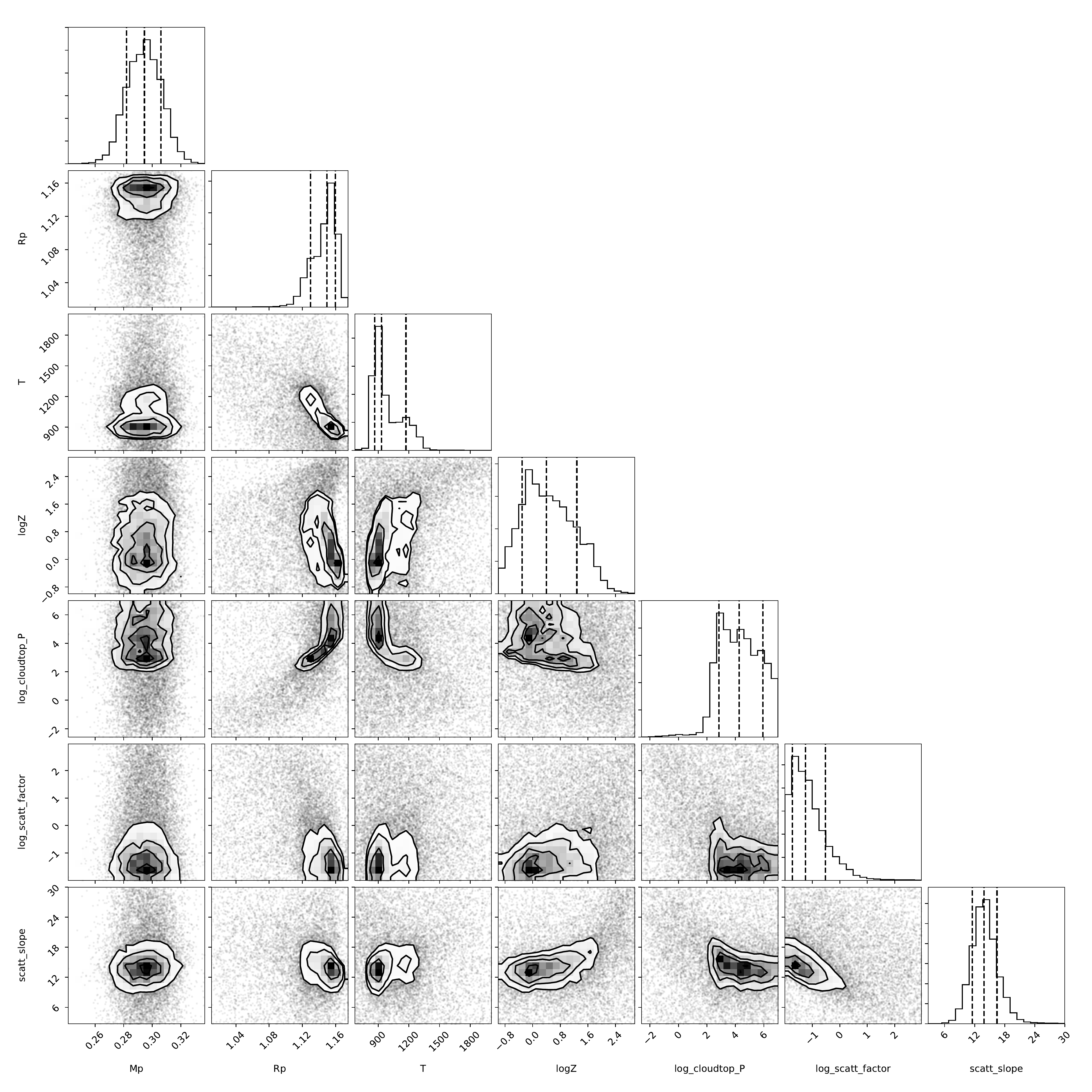}
    \caption{Corner plot of fitted parameters for the best fitting atmospheric model obtained using PLATON (Model 3, Table \ref{tab:3night_table}). Vertical dashed lines show the median and the 16\% and 84\% quartiles.}
    \label{fig:corner_plot}
\end{figure*}

\section{Additional Retrieval Results}

\begin{table*}
\centering
\caption{PLATON retrieval results for the parameters as defined in Table \ref{tab:platon_priors} for the transmission spectrum of Night 1.}
\label{tab:n1_table}
\begin{tabular}{ccccccccc}
\cline{2-9}
 & \multicolumn{2}{c|}{Model 1} & \multicolumn{2}{c|}{Model 2} & \multicolumn{2}{c|}{Model 3} & \multicolumn{2}{|c}{Model 4} \\ \cline{2-9} 
 & Median & Best Fit & Median & Best Fit & Median & Best Fit & Median & Best Fit \\ \hline
\multicolumn{1}{|l|}{$\ln \chi$} &  & 115.0 &  & 112.6 &  & 115.0 &  & 108.1 \\ \hline
\multicolumn{1}{|l|}{$R_\mathrm{S}$ ($R_{\odot} $)} & 1.186 (fixed) & - & - & - & - & - & - & - \\
\multicolumn{1}{|l|}{$T_\mathrm{eff}$ (K)} & 5800 (fixed) & - & - & - & - & - & - & - \\
\multicolumn{1}{|l|}{$M_\mathrm{P}$ ($M_\mathrm{J}$)} & $0.29^{+0.01}_{-0.01}$ & 0.30 & $0.30^{+0.01}_{-0.01}$ & 0.29 & $0.29^{+0.01}_{-0.01}$ & 0.29 & $0.29 ^{+0.01}_{-0.01}$ & 0.29 \\
\multicolumn{1}{|l|}{$R_\mathrm{P}$ ($R_\mathrm{J}$)} & $1.10^{+0.04}_{-0.05}$ & 1.14 & $1.07^{+0.02}_{-0.03}$ & 1.10 & $1.12^{+0.03}_{-0.05}$ & 1.16 & $1.12^{+0.01}_{-0.01}$ & 1.13 \\
\multicolumn{1}{|l|}{$T_\mathrm{limb}$ (K)} & $1049^{+289}_{-211}$ & 892 & $923^{+306}_{-149}$ & 897 & $1082^{+247}_{-224}$ & 892 & $1201^{+88}_{-159}$ & 1291 \\
\multicolumn{1}{|l|}{$\log Z$} & $1.44^{+0.85}_{-1.34}$ & 0.30 & $1.20^{+1.28}_{-1.42}$ & 0.61 & $1.28^{+0.78}_{-1.23}$ & 0.06 & $0.06^{+0.83}_{-0.68}$ & -0.84 \\
\multicolumn{1}{|l|}{$\log P_\mathrm{cloud}$ (Pa)} & $2.35^{+2.91}_{-2.79}$ & 6.89 & $3.26^{+2.48}_{-2.95}$ & 5.30 & $1.74^{+2.42}_{-2.35}$ & 4.32 & $4.49^{+1.53}_{-1.40}$ & 3.73 \\
\multicolumn{1}{|l|}{$\log s$} & $-0.59^{+1.91}_{-0.99}$ & -1.85 & $0.48^{+1.37}_{-1.40}$ & 0.32 & $-0.36^{+1.73}_{-1.09}$ & -1.72 & $1.93^{+0.59}_{-0.57}$ & 1.01 \\
\multicolumn{1}{|l|}{$\alpha$} & $17.3^{+6.5}_{-7.1}$ & 14.4 & 4 (fixed) & - & $18.7^{+6.0}_{-4.8}$ & 15.3 & 4 (fixed) & - \\
\multicolumn{1}{|l|}{$T_\mathrm{spot}$ (K)} & $5530^{+344}_{-690}$ & 5239 & $5280^{+241}_{-499}$ & 5554 & $T_\mathrm{eff}$ (fixed) & - & - & - \\
\multicolumn{1}{|l|}{$f_\mathrm{active}$} & $0.28^{+0.41}_{-0.18}$ & 0.13 & $0.40^{+0.27}_{-0.15}$ & 0.65 & 0 (fixed) & - & - & - \\
\multicolumn{1}{|l|}{C/O} & 0.53 (fixed) & - & - & - & - & - & - & - \\ \hline
\end{tabular}
\end{table*}

\begin{table*}
\centering
\caption{PLATON retrieval results for the parameters as defined in Table \ref{tab:platon_priors} for the transmission spectrum of Night 2.}
\label{tab:n2_table}
\begin{tabular}{ccccccccc}
\cline{2-9}
 & \multicolumn{2}{c|}{Model 1} & \multicolumn{2}{c|}{Model 2} & \multicolumn{2}{c|}{Model 3} & \multicolumn{2}{|c}{Model 4} \\ \cline{2-9} 
 & Median & Best Fit & Median & Best Fit & Median & Best Fit & Median & Best Fit \\ \hline
\multicolumn{1}{|l|}{$\ln \chi$} &  & 113.1 &  & 113.5 &  & 111.7 &  & 110.5 \\ \hline
\multicolumn{1}{|l|}{$R_\mathrm{S}$ ($R_{\odot} $)} & 1.186 (fixed) & - & - & - & - & - & - & - \\
\multicolumn{1}{|l|}{$T_\mathrm{eff}$ (K)} & 5800 (fixed) & - & - & - & - & - & - & - \\
\multicolumn{1}{|l|}{$M_\mathrm{P}$ ($M_\mathrm{J}$)} & $0.29^{+0.01}_{-0.01}$ & 0.30 & $0.30^{+0.01}_{-0.01}$ & 0.30 & $0.29^{+0.01}_{-0.01}$ & 0.29 & $0.29^{+0.01}_{-0.01}$ & 0.29 \\
\multicolumn{1}{|l|}{$R_\mathrm{P}$ ($R_\mathrm{J}$)} & $1.10^{+0.04}_{-0.04}$ & 1.03 & $1.10^{+0.03}_{-0.03}$ & 1.04 & $1.14^{+0.02}_{-0.02}$ & 1.16 & $1.14^{+0.02}_{-0.02}$ & 1.16 \\
\multicolumn{1}{|l|}{$T_\mathrm{limb}$ (K)} & $1091^{+276}_{-199}$ & 853 & $986^{+287}_{-123}$ & 871 & $1120^{+237}_{-205}$ & 902 & $1141^{+196}_{-205}$ & 892 \\
\multicolumn{1}{|l|}{$\log Z$} & $1.11^{+1.04}_{-1.30}$ & 0.58 & $1.05^{+0.71}_{-1.00}$ & 0.52 & $1.19^{+1.00}_{-0.92}$ & 0.02 & $0.48^{+0.86}_{-0.91}$ & -0.39 \\
\multicolumn{1}{|l|}{$\log P_\mathrm{cloud}$ (Pa)} & $3.63^{+2.05}_{-1.43}$ & 5.86 & $4.17^{+1.75}_{-1.66}$ & 4.65 & $3.62^{+2.00}_{-1.62}$ & 4.43 & $4.44^{+1.61}_{-1.40}$ & 6.54 \\
\multicolumn{1}{|l|}{$\log s$} & $-0.04^{+1.80}_{-1.30}$ & -0.32  & $0.24^{+1.17}_{-1.34}$ & -1.48 & $-0.39^{+1.46}_{-1.04}$ & -1.06 & $1.27^{+0.68}_{-0.77}$ & 0.31 \\
\multicolumn{1}{|l|}{$\alpha$} & $4.9^{+7.4}_{-5.6}$ & 4.0 & 4 (fixed) & - & $10.8^{+4.9}_{-5.1}$ & 10.0 & 4 (fixed) & - \\
\multicolumn{1}{|l|}{$T_\mathrm{spot}$ (K)} & $5356^{+323}_{-808}$ & 3960 & $5354^{+295}_{-973}$ & 4095 & $T_\mathrm{eff}$ (fixed) & - & - & - \\
\multicolumn{1}{|l|}{$f_\mathrm{active}$} & $0.25^{+0.28}_{-0.12}$ & 0.27 & $0.26^{+0.41}_{-0.12}$ & 0.27 & 0 (fixed) & - & - & - \\
\multicolumn{1}{|l|}{C/O} & 0.53 (fixed) & - & - & - & - & - & - & - \\ \hline
\end{tabular}
\end{table*}

\begin{table*}
\centering
\caption{PLATON retrieval results for the parameters as defined in Table \ref{tab:platon_priors} for the transmission spectrum of Night 3.}
\label{tab:n3_table}
\begin{tabular}{ccccccccc}
\cline{2-9}
 & \multicolumn{2}{c|}{Model 1} & \multicolumn{2}{c|}{Model 2} & \multicolumn{2}{c|}{Model 3} & \multicolumn{2}{|c}{Model 4} \\ \cline{2-9} 
 & Median & Best Fit & Median & Best Fit & Median & Best Fit & Median & Best Fit \\ \hline
\multicolumn{1}{|l|}{$\ln \chi$} &  & 117.1 &  & 116.5 &  & 115.9 &  & 107.3 \\ \hline
\multicolumn{1}{|l|}{$R_\mathrm{S}$ ($R_{\odot} $)} & 1.186 (fixed) & - & - & - & - & - & - & - \\ 
\multicolumn{1}{|l|}{$T_\mathrm{eff}$ (K)} & 5800 (fixed) & - & - & - & - & - & - & - \\
\multicolumn{1}{|l|}{$M_\mathrm{P}$ ($M_\mathrm{J}$)} & $0.29^{+0.01}_{-0.01}$ & 0.29 & $0.30^{+0.01}_{-0.01}$ & 0.29 & $0.29^{+0.01}_{-0.01}$ & 0.30 & $0.29^{+0.01}_{-0.01}$ & 0.29 \\
\multicolumn{1}{|l|}{$R_\mathrm{P}$ ($R_\mathrm{J}$)} & $1.13^{+0.04}_{-0.08}$ & 1.04 & $1.05^{+0.04}_{-0.03}$ & 1.02 & $1.13^{+0.01}_{-0.01}$ & 1.14 & $1.10^{+0.02}_{-0.01}$ & 1.10 \\
\multicolumn{1}{|l|}{$T_\mathrm{limb}$ (K)} & $830^{+68}_{-95}$ & 832 & $878^{+192}_{-70}$ & 807 & $846^{+69}_{-94}$ & 828 & $1170^{+96}_{-176}$ & 1219 \\
\multicolumn{1}{|l|}{$\log Z$} & $0.47^{+0.91}_{-0.61}$ & 0.22 & $0.65^{+1.12}_{-0.84}$ & -0.21 & $0.53^{+0.94}_{-0.75}$ & -0.36 & $0.29^{+0.63}_{-0.73}$ & 0.32 \\
\multicolumn{1}{|l|}{$\log P_\mathrm{cloud}$ (Pa)} & $5.31^{+1.01}_{-1.00}$ & 6.01 & $5.13^{+1.14}_{-1.21}$ & 6.24 & $5.32^{+1.02}_{-1.07}$ & 6.52 & $4.79^{+1.32}_{-1.26}$ & 3.51 \\
\multicolumn{1}{|l|}{$\log s$} & $-0.74^{+0.67}_{-0.63}$ & -0.40 & $0.15^{+0.84}_{-0.83}$ & -0.73 & $-0.41^{+0.70}_{-0.60}$ & -0.95 & $1.66^{+0.54}_{-0.54}$ & 1.64 \\
\multicolumn{1}{|l|}{$\alpha$} & $13.0^{+8.5}_{-5.8}$ & 7.4 & 4 (fixed) & - & $11.9^{+3.4}_{-2.3}$ & 11.4 & 4 (fixed) & - \\
\multicolumn{1}{|l|}{$T_\mathrm{spot}$ (K)} & $5829^{+242}_{-1851}$ & 3844 & $4310^{+1099}_{-331}$ & 4066 & $T_\mathrm{eff}$ (fixed) & - & - & - \\
\multicolumn{1}{|l|}{$f_\mathrm{active}$} & $0.30^{+0.45}_{-0.17}$ & 0.19 & $0.23^{+0.13}_{-0.05}$ & 0.26 & 0 (fixed) & - & - & - \\
\multicolumn{1}{|l|}{C/O} & 0.53 (fixed) & - & - & - & - & - & - & - \\ \hline
\end{tabular}
\end{table*}

\bsp	
\label{lastpage}
\end{document}